\documentclass[12pt,preprint]{aastex}

\shorttitle{Terrestrial Formation and Type I Migration}
\shortauthors{McNeil et al.}

\begin{document}


\title{Effects of Type I Migration on Terrestrial Planet Formation}


\author{Douglas McNeil and Martin Duncan}
\affil{Department of Physics and Astronomy, Queen's University,\\
  Kingston, Ontario, Canada K7L 3N6} \email{mcneil@astro.queensu.ca}

\author{Harold F. Levison} \affil{Southwest Research Institute, Boulder,
  Colorado, USA 23489}


\begin{abstract}
Planetary embryos embedded in a gas disc suffer a decay in semimajor
axis -- type I migration -- due to the asymmetric torques produced by
the interior and exterior wakes raised by the body \citep{goldtrem,
ward}.  This presents a challenge for standard oligarchic approaches
to forming the terrestrial planets \citep{ko2} as the timescale to
grow the progenitor objects near 1 AU is longer than that for them to
decay into the Sun.  In this paper we investigate the middle and late
stages of oligarchic growth using both semi-analytic methods (based
upon \citealt{edolig}) and N-body integrations, and vary gas
properties such as dissipation timescale in different models of the
protoplanetary disc. We conclude that even for near-nominal migration
efficiencies and gas dissipation timescales of $\sim\!1$ Myr it is
possible to maintain sufficient mass in the terrestrial region to form
Earth and Venus if the disc mass is enhanced by factors of $\sim2-4$
over the minimum mass model. The resulting configurations differ in
several ways from the initial conditions used in previous simulations
of the final stages of terrestrial accretion (e.g.~\citealt{cham2}),
chiefly in (1) larger inter-embryo spacings, (2) larger embryo masses,
and (3) up to $\sim\!0.4 M_\earth$ of material left in the form of
planetesimals when the gas vanishes. The systems we produce are
reasonably stable for $\sim\!100$ Myr and therefore require an
external source to stir up the embryos sufficiently to produce final
systems resembling the terrestrial planets.
\end{abstract}




\section{Introduction}

A standard model for forming the terrestrial planets has emerged, and
is divided broadly into three stages (see references in
\citealt*{canupbook}). The first stage is poorly understood, but the
formation of the Sun is thought to leave a large but thin
protoplanetary disc of both gas and solids, with coagulation or
gravitational instability producing rocky objects (`planetesimals') in
the metre-to-km size range.  In the second stage, interactions between
the planetesimals result in an early phase of `runaway growth', to be
discussed below, where the single largest body in a given region
breaks away from the mass spectrum and becomes significantly larger
than the second-largest.  Over $\sim10^5 - 10^6$ years, this gives
rise to an oligarchic phase, also discussed below, which produces many
well-spaced comparable-mass embryos.  In the third stage, the embryos
interact with each other and what remains of the the disc, merging and
producing the terrestrial planets we observe today on a timescale of
$\sim10^7 - 10^8$ years.

 The basic paradigm for the second stage, the `planetesimal problem',
is described in \citet{weth93}.  One considers a very large number of
small bodies of comparable mass, moving in low-eccentricity,
low-inclination orbits around the Sun.  These objects will suffer
close encounters with each other, changing the velocity distribution,
and occasionally collide and either merge or fragment, changing the
mass spectrum as well.  Energy equipartition in the system decreases
the random velocities of the larger bodies, and increases the
velocities of the smaller ones, an effect known as `dynamical
friction'.  It has been shown \citep{weth89} that dynamical friction
leads to runaway growth of the largest body, as the decrease in
velocity increases the collisional cross section and thus the
accretion rate.  Later simulations \citep*[e.g.][]{ida92a, ida92b,
ko1, ko2} confirmed this, and demonstrated that the end result of
runaway growth was an `oligarchic' phase where the accretion process
is dominated by the gravitational stirring of well-separated embryos
of roughly equal masses.

All simulations published to date of the second stage have either
neglected any interaction between the gas disk and the solid bodies or
have included only the aerodynamic gas drag acting on the solid bodies
as described by \citet{adachi}.  However, it is well known that the
interactions between the protoplanetary disc and the protoplanets can
also cause the latter to migrate.  This can happen in several ways:
type I, where the migration is driven by the asymmetry between the
torques generated by the interior and exterior wakes of the body
\citep{goldtrem, ward}; type II, where the protoplanet is large enough
to open up a gap in the disc, coupling it to the viscous evolution of
the disc \citep{papandlin}; and recently type III \citep{masset,
artym}, a very fast migration mode which occurs when the embryo begins
to migrate quickly enough that corotating material cannot librate.
(The solid material in the disc can also be scattered by the embryos,
leading to embryo migration directly; this `type 0' migration
\citep{fernip} is thought to play an important role in the evolution
of the outer solar system.)

In most models, protoplanets of terrestrial mass are incapable of
opening a gap and therefore do not suffer type II migration, and
likewise type III should not be relevant \citep{masset}.  Accordingly
we restrict ourselves to considering the effects of type I migration,
specifically the tidal interactions between the planet and disc which
result in semimajor axis decay and random velocity decay (both
eccentricity and inclination damping.)  In the models we are
considering, the typical effect is to decrease the semimajor axis,
which leads to difficulties for accretion: the type I migration rate
scales with the mass of the object, so as an object grows it moves
inwards faster.  Indeed, the type I migration timescales for the
embryos produced near 1 AU are typically comparable to the timescales
for their growth; this problem is discussed in \cite{tanakaward}.

In a sense, we seek to refine the initial conditions for the third
stage, by including the effects of the tidal interaction during the
second stage.  This is very similar in spirit to \cite{komin}, but
they include only the effects of random velocity damping, not of
semimajor axis decay, and study only the third stage itself.  To this
end we will investigate the effects of different assumptions about the
disc and the migration efficiency on the state of the protoplanetary
system when the gas vanishes, and thus our ability to produce
terrestrial planets from what material remains.  Note that we
concentrate here on forming the Earth and Venus.  Mercury has a small
enough mass ($0.055 \, M_\earth$) that it is easily considered debris
at this scale, and the problem of forming Mars has its own special
difficulties involving the asteroid belt, the possible presence of
other large embryos, and influences from Jupiter, which we defer.  We
will not discuss here the third stage resulting from this work,
because it is probably dependent on the details of the stirring of the
embryos near the end of the second stage. This is likely to involve
the perturbations introduced by the formation of Jupiter and Saturn,
for which models are sufficiently uncertain that we reserve a
discussion of these effects for a subsequent paper.

In \S\ref{section:model} we develop a semi-analytic model for oligarchy
during type I migration in the terrestrial region and use the results
to motivate the N-body simulations which follow.  In
\S\ref{section:nbody_methods} we discuss our numerical N-body methods,
and present the results of our N-body simulations in
\S\ref{section:sims}.  Section \ref{section:disc} contains a further
discussion of our scenario and we summarize our conclusions in
\S\ref{section:conc}.

\section{Models}
\label{section:model}

In this section we construct a model of oligarchic formation by
adapting the model of \cite{edolig}, which built on the earlier work
of \cite{ko2, ko2000, ko_div}.  We content ourselves with summarizing
their results and noting our modifications.  We then discuss the
differences between the physics in this semi-analytic model and the
physics we use in our N-body simulations.

We use the semi-analytic model for two distinct purposes: first, to
provide a quick way to explore parameter space and determine what
regimes are good candidates for more detailed N-body investigation;
and second, to generate initial conditions for said N-body
simulations.  This allows us to begin our runs later in the process
than would otherwise be possible, thereby avoiding the expense of
handling very large numbers of embryos at early times.

\subsection{Semi-analytic Model}

\cite{edolig} represent the oligarchic system by a two-component model
consisting of embryos and field planetesimals, where the former
accrete but the latter do not.  The model variables are the embryo
mass $M$ and planetesimal surface density $\Sigma_{m}$ as functions of
semimajor axis $a$, and we seek to evolve $M(a)$ and $\Sigma_{m}(a)$
over time $t$.

Note that since $M(a)$ is continuous but is meant to represent a
discrete population, it corresponds not to the mass of material in the
form of embryos in a given region, but to the mass that an embryo
would have were one present at that semimajor axis.  It is also
assumed that the embryo separation $b$ measured in single-planet Hill
units (defined by $r_H = (M/3 M_\Sun)^{1/3} a$) is constant and
contains all the information about the effects of any embryo-embryo
mergers in the model.  This assumption is justified in \cite{ko2} as
the consequence of an equilibration between the decrease in $b$ caused
by accretion of small objects and an increase in $b$ caused by
two-body scattering followed by eccentricity damping due to dynamical
friction from the field of planetesimals.  The properties of the
resulting (non-continuous) embryo population can be determined from
$M(a)$ by discretizing the curve appropriately.  Thus we construct an
effective embryo surface density $\Sigma_M = M / (2 \pi \, a \, \Delta
a)$, setting the embryo spacing $\Delta a = b \, r_H$.

The embryos are assumed to have uniform physical density $\rho_M$.  It
is also assumed that the embryos are kept in near-circular orbits by
dynamical friction with the field (or, when active, tidal damping from
the gas) and therefore the important random velocity is that of the
field particles.

The field population is similarly approximated as being comprised of
planetesimals of uniform mass $m$ and density $\rho_m$ (and therefore
uniform radius $r_m$) with (root-mean-square) eccentricity $e_m$ and
inclination $i_m$ satisfying $e_m = 2 i_m$.  \cite{edolig} determine
the eccentricity, after \cite{ida93}, by equating the
stirring timescale of the field by the embryos with the damping
timescale due to aerodynamic drag on the field particles.

The stirring timescale in a dispersion-dominated regime can be written
as

\begin{equation}
\label{eq:tstir}
T_{\rm{stir}} \simeq \frac{1}{40} \left( \frac{\Omega^2 a^3}{G M}
  \right)^2 \frac{e_m^4 \, M}{\Sigma_M a^2 \Omega},
\end{equation}

where $G$ is the gravitational constant and $\Omega$ the orbital
frequency.  We use the damping timescale due to aerodynamic drag from
\cite{adachi}, where $\rho_{\rm{gas}}$ is the density of the gas and
$C_D$ is a drag efficiency factor ($\simeq 1$) dependent upon the
Reynolds number of the disc.  We neglect resonant interactions.  In
our notation,

\begin{equation}
\label{eq:tdamp}
T_{\rm{aero}} = \frac{1}{e_m^2} \frac{m}{(C_D/2) \, \pi r_m^2 \,
  \rho_{\rm{gas}} \, a \, \Omega}
.
\end{equation}

Equating these two timescales yields an equilibrium eccentricity of
\begin{equation}
\label{eq:eccequil}
e_{\rm{eq}} \simeq \frac {1.72 \,\, m^{1/15} \, M^{1/3} \,
\rho_{\rm{m}}^{2/15}} {{M_\Sun}^{1/3}\, C_D^{1/5}\,
\rho_{\rm{gas}}^{1/5}\, a^{1/5} \, b^{1/5}}
.
\end{equation}

\cite{edolig} derive an expression for the growth of the embryo mass
\begin{equation}
\label{eq:dMdt}
\frac{dM}{dt} = \frac{3.93 \, M_\Sun^{1/6} \, G^{1/2} \,
\Sigma_{\rm{m}} \, M^{2/3} \, C_D^{2/5} \, \rho_{\rm{gas}}^{2/5}}
{\rho_{M}^{1/3} \, a^{1/10} \, m^{2/15} \, \rho_{m}^{4/15}}
,
\end{equation}

and a corresponding decrease in planetesimal surface density from conservation of mass 
\begin{equation}
\label{eq:dsdt}
\frac{d\Sigma_{m}}{dt} = \frac{ -M_\Sun^{1/3} }{3^{2/3} \, b \, \pi a^2
  \, M^{1/3}} \frac{dM}{dt}
.
\end{equation}

(The aerodynamic drag also affects $\Sigma_m$, as we discuss later.)

We restrict consideration of initial disc conditions at $T=0$ to those
resembling the minimum mass model \citep{hay1981}.  Specifically, we
assume the surface density in solids has the form $\Sigma_{\rm{solid}}
= k (r/{\rm AU})^{p}$; the nominal values are $k = 7.1 {\rm \, g
\,cm^{-2}}, p = -1.5$ and we introduce an enhancement factor $f_{\rm
enh}$ so that $\Sigma_{\rm{solid}} = f_{\rm enh} \, (7.1 {\rm
\,g\,cm^{-2}}) \, (r/{\rm AU})^{p}$.  We adopt a gas-to-solid ratio of
240, so that $\Sigma_{\rm{gas}} = 240 \, \Sigma_{\rm{solid}}$. We
assume the disc has an exponential vertical structure such that
$\rho_{\rm{gas}}(r,z) = \rho_{\rm{gas}}(r) \exp(-z^2/z_0^2)$, with
$\rho_{\rm{gas}}(r)$ taken at the midplane and $z_0(r)$ the disc
half-thickness.  (Integrating the z-dependent term over all $z$ gives
the relation $\rho_{\rm{gas}} = \Sigma_{\rm{gas}} / \sqrt{\pi} z_0$.)
The half-thickness varies as $z_0(r) = Z_{1} (r/\rm{AU})^{5/4}$, where
we take the thickness at 1 AU to be $Z_{1} = \rm{0.07 \, AU}$.  We
model the dissipation of the gas by introducing an exponential decay
with a characteristic time $\tau_{\rm{decay}}$ which is fixed for a
given model and does not vary through the disc, $\rho \propto
\exp(-t/\tau_{\rm{decay}})$, where we allow $\tau_{\rm{decay}} =
\infty$.  We define a parameter $\eta$ which measures the degree of
pressure support of the disc (and thus deviations of circular orbits
from Keplerian velocity): $\eta = (\pi/16) (\alpha + \beta) (z_0/a)^2$
where $\alpha = 5/4 - p$ and $\beta = 1/2$; typically $\eta \simeq
0.001$.  We assume the gas is in cylindrical rotation such that
$v_{\rm{gas}} = v_{\rm{kep}} \sqrt{1 - 2 \eta}$.

For the planetesimal migration introduced by aerodynamic drag, we
 adopt the orbit-averaged approximation valid for small $e$, $i$, and
 $\eta$ due to \cite{adachi}:

\begin{equation}
\label{eq:v_aero}
v_m = \frac{da}{dt} \! \Big\arrowvert_{\!\rm{aero}} \simeq
- 2 \frac{a}{T_{\rm{aero}} \, e_m} \left( \frac{5}{8} e_m^2 +
  \frac{1}{2} i_m^2 + \eta^2 \right)^{1/2} 
  \bigg\{ \eta + \left( \frac{\alpha}{4} + \frac{5}{16} \right) e_m^2 + \frac{1}{8}
    i_m^2 \bigg\}
.
\end{equation}

We also add a prescription for type I migration based on that of
\cite{paplar}, assuming that the embryos are on circular orbits (see
also \citealt{tanakaward}). They derive a timescale for semimajor axis
damping $t_a$:

\begin{equation}
\label{eq:ta_semianal}
  t_a = \frac{1}{c_a} \, 
  \left(\frac{a^3}{G M_\Sun}\right)^{1/2}
  \left(\frac{Z_1}{{\rm{AU}}} \, {\left(\frac{a}{\rm{AU}}\right)}^{1/4} \right)^{2} 
  \left(\frac{\Sigma_{\rm{gas}} \pi a^2}{M_\Sun}\right)^{-1} \,
  \left(\frac{M}{M_\Sun}\right)^{-1} \,
,
\end{equation}
where $c_a$ is a migration efficiency parameter, with $c_a = 1$ being
nominal and $c_a = 0$ yielding $t_a = \infty$.  By varying $c_a$ in
our simulations we can account for an uncertainty of order unity in
the rate of migration.

For the purpose of this model we can construct a rough radial velocity
for protoplanet material

\begin{equation}
\label{eq:v_M}
v_M = \frac{da}{dt} \! \Big\arrowvert_{\!\rm{tidal}} = -\frac{a}{t_a}
\end{equation}

We emphasize that since it is not clear (to choose one assumption
among many) whether the aerodynamic drag coefficient $C_D$ should be
$\simeq$1 or $\simeq$2 \citep{adachi}, the above model is only
expected to be accurate to within factors of order unity.  This is
true even if we neglect consequences of assuming that all field
particles have the same fixed mass and the issue of the evolution of
inter-embryo spacing (see \S\ref{bevol}).

\begin{figure}
\epsscale{.80}
\plotone{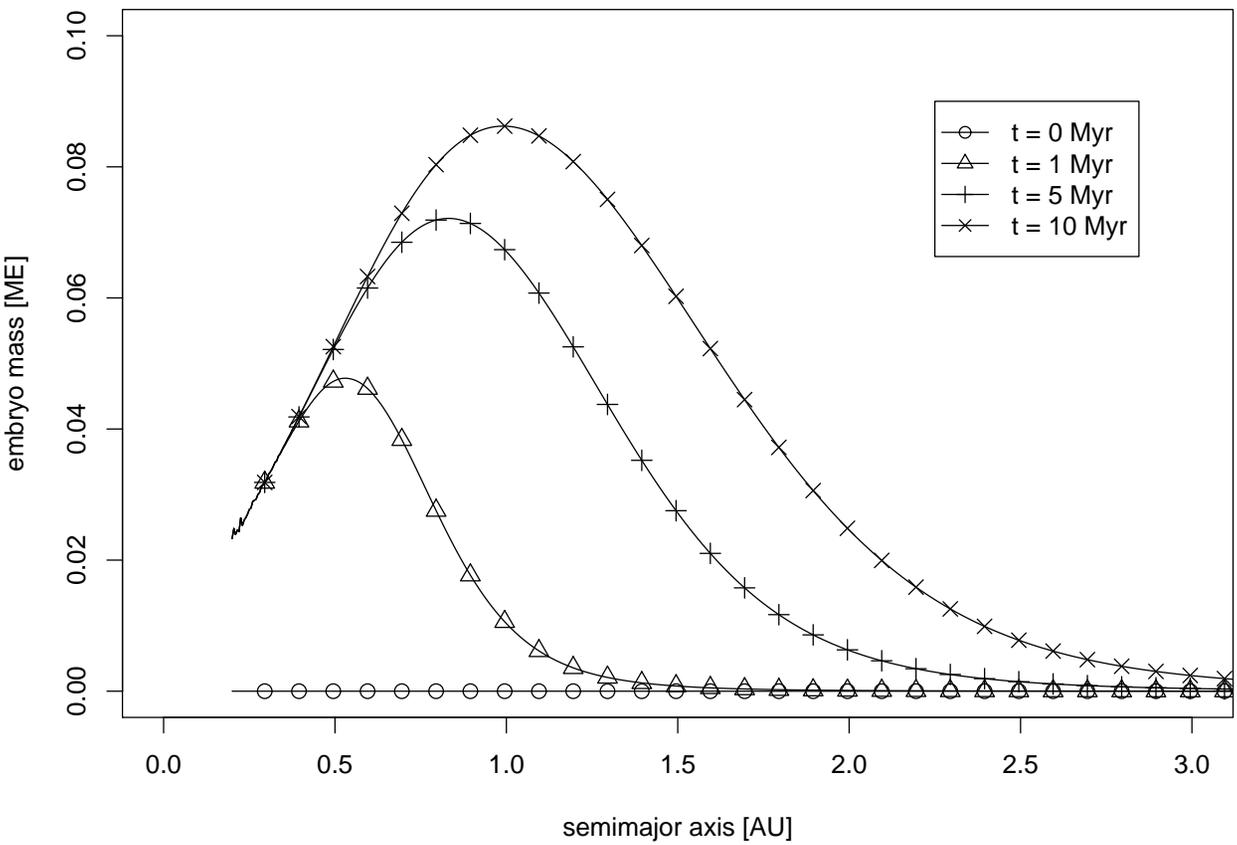}
\caption{Evolution of nominal model without tidal migration, $k = 7.1
  \rm{\,g \, cm^{-2}}$, $p = -1.5$, $\tau_{\rm{decay}} = \infty$.
\label{figure:ed_1}
}

\end{figure}

\subsection{Results from Semi-analytic Model}

We now integrate the model described in the previous section forward
in time.  We have two variables, the embryo mass $M(a)$ and the
planetesimal surface density $\Sigma_m(a)$, whose evolution (in the
absence of aerodynamic or tidal migration) is specified by equations
(\ref{eq:dMdt}) and (\ref{eq:dsdt}).  We introduce migration through
equations (\ref{eq:v_aero}), (\ref{eq:ta_semianal}), and
(\ref{eq:v_M}).  These five equations are simultaneously integrated
across a zone from $a =$ 0.2 AU to $a =$ 5.0 AU, subject to our
time-dependent gas profile.

Figure \ref{figure:ed_1} shows the evolution of the standard
minimum-mass model (k = 7.1 $\rm{g \, cm^{-2}}$, $f_{\rm enh} = 1.0$,
p = -1.5, $\tau_{\rm{decay}} = \infty$) assuming that $b = 10$.  We
arbitrarily set $M_0 = 10^{-6} M_\earth$ as the seed mass for the
embryos at all $a$, so that the total fraction of mass in embryos is
$\lesssim 5 \times 10^{-4}$ that of the mass in the planetesimal field;
accretion at early times is so quick that the precise value is
unimportant as long as $M_0 \ll M_{\rm{final}}$.  As is evident on the
left part of the curves, the embryo mass nearly asymptotes toward a
power-law in semimajor axis $a$: except for deviations introduced by
planetesimal migration and the nonzero initial mass $M_0$ of the
embryo, the limiting mass is set solely by the amount of planetesimal
material and the chosen $b$, $M_{\rm lim} = \sqrt{8/(3 M_\sun)}
(\Sigma_m \, \pi \, b \, a^2)^{3/2}$.  (The right side of the curve
will reach the same asymptote eventually.)

\begin{figure}
\epsscale{.80}
\plotone{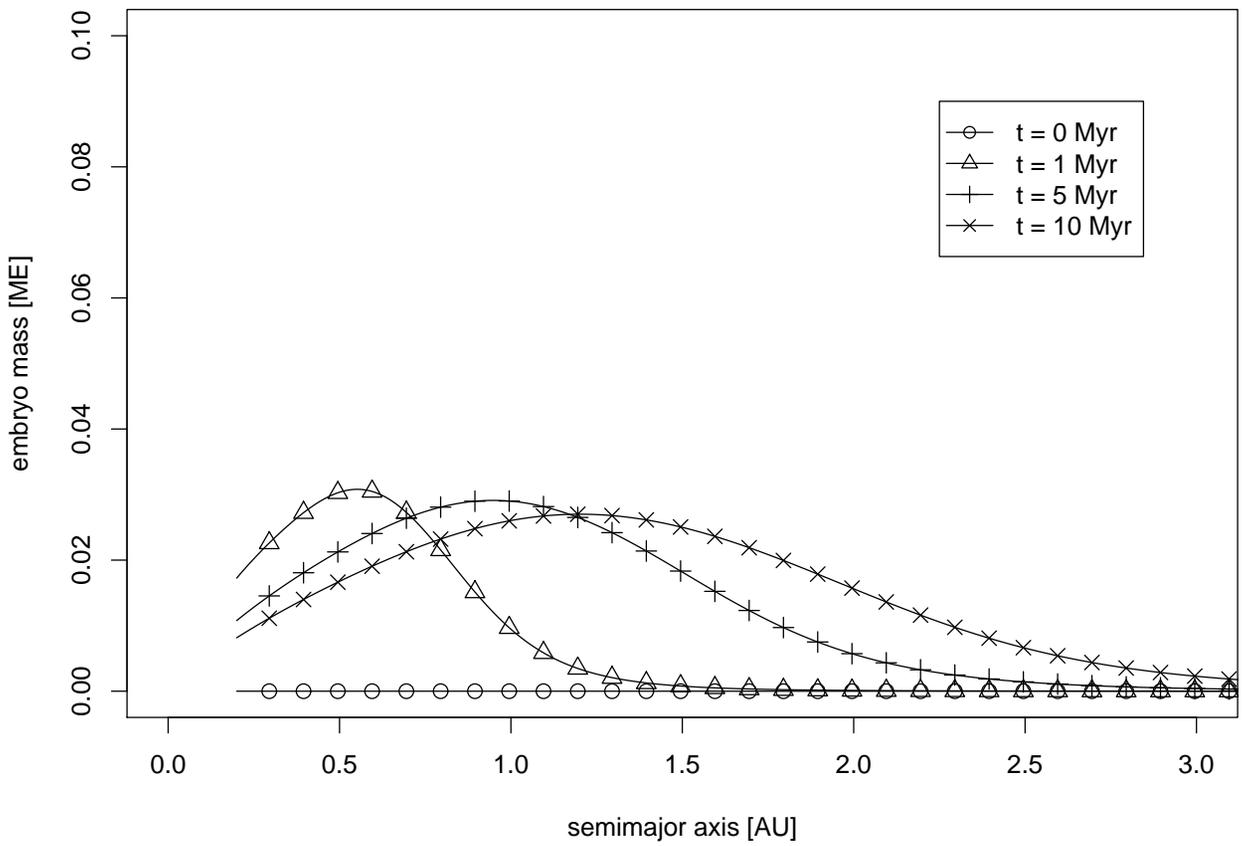}
\caption{Evolution of nominal model with migration, $k = 7.1 \rm{\,g \,
    cm^{-2}}$, $p = -1.5$, $\tau_{\rm{decay}} = \infty$.
\label{figure:ed_2}
}
\end{figure}

Figure \ref{figure:ed_2} shows the minimum-mass model with tidal
migration added (keeping $\tau_{\rm{decay}} = \infty$): several
differences from the previous case are immediately apparent.  The
maximum embryo mass is substantially reduced, from $0.086 M_\earth$ at
10 Myr to $0.027 M_\earth$.  Without migration there is a sharp
drop-off in embryo mass past the peak of the accretion front, but with
tidal migration there is a wide band in semimajor axis where the
embryo mass is within 25\% of the maximum.  With $\tau_{\rm{decay}} =
\infty$ there is nothing to prevent embryo material from continuing to
migrate inwards as it accretes, and thus eventually all embryo mass
falls into the Sun.

\begin{figure}
\epsscale{.80}
\plotone{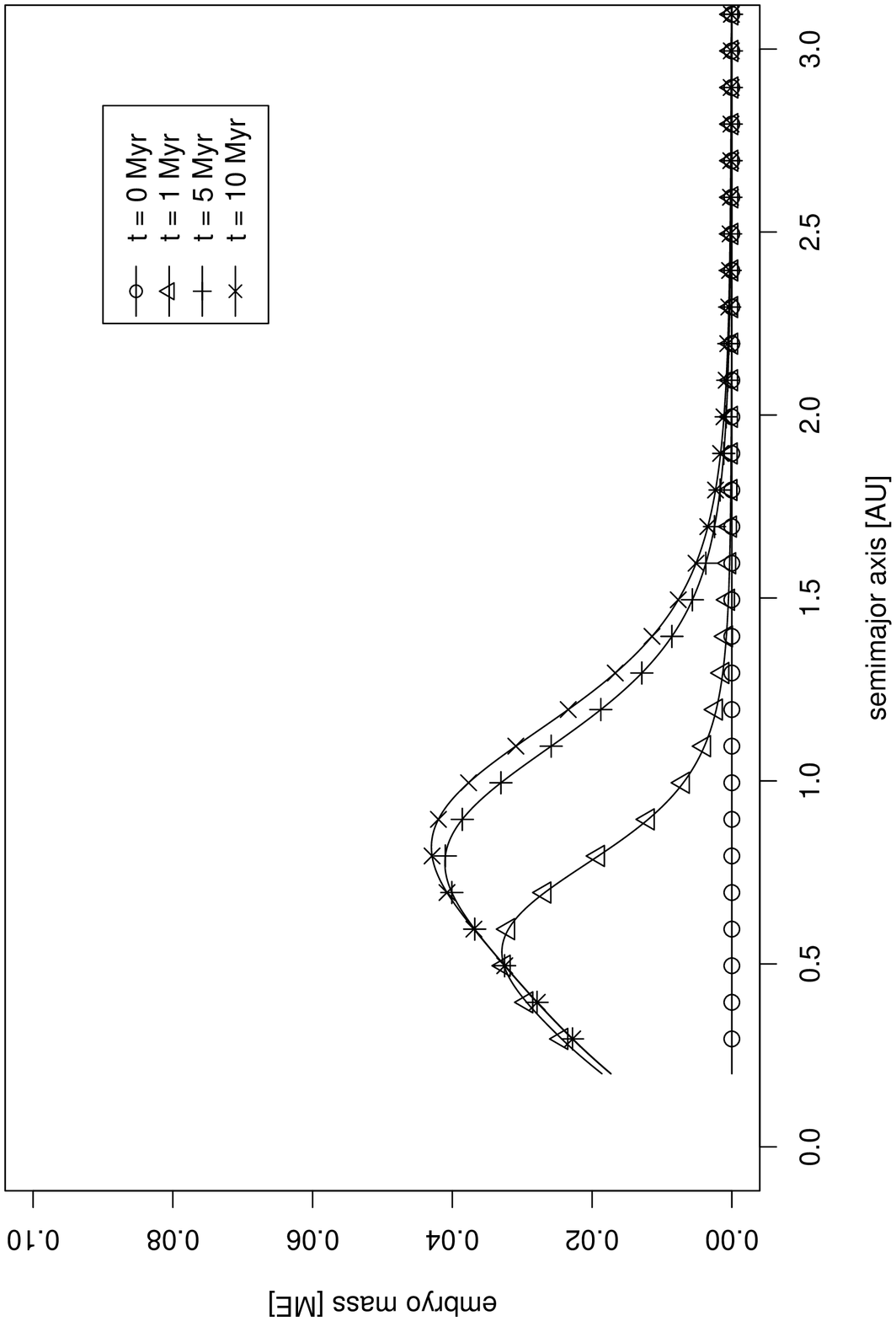}
\caption{Evolution of nominal model with migration and dissipation, $k = 7.1
  \rm{\,g \, cm^{-2}}$, $p = -1.5$, $\tau_{\rm{decay}}$ = 1 Myr.
\label{figure:ed_3}
}
\end{figure}

Figure \ref{figure:ed_3} is the result of introducing gas dissipation
of $\tau_{\rm{decay}}$ = 1 Myr, which produces an intermediate system.
The maximum embryo mass reached is $\sim\!0.043 M_\earth$, down by a
factor of 2 from the no-migration case, but the gas decay has several
effects: type I migration is halted, preserving embryo mass; the
increase in planetesimal eccentricity as a result of the decrease in
aerodynamic drag increases the accretion time; and as a result of this
increase, a fair amount of material is left in the form of
planetesimals.  However, the total amount of material in the
terrestrial region ($0.5 {\rm AU} \leq a \leq 1.5 {\rm AU}$) is
insufficient to later accrete into the Earth and Venus
during stage three.

The dissipation timescale of $\sim1$ Myr was chosen based on studies
of the disc lifetimes around stars in young clusters by
\citealt*{haisch}.  They find that roughly half the stars lose their
discs in $\lesssim3$ Myr, and the overall disc lifetime was $\sim6$
Myr.  (In the N-body simulations to be discussed later, we will
consider discs with both larger and greater $\tau_{\rm{decay}}$.)

\begin{figure}
\epsscale{.80}
\plotone{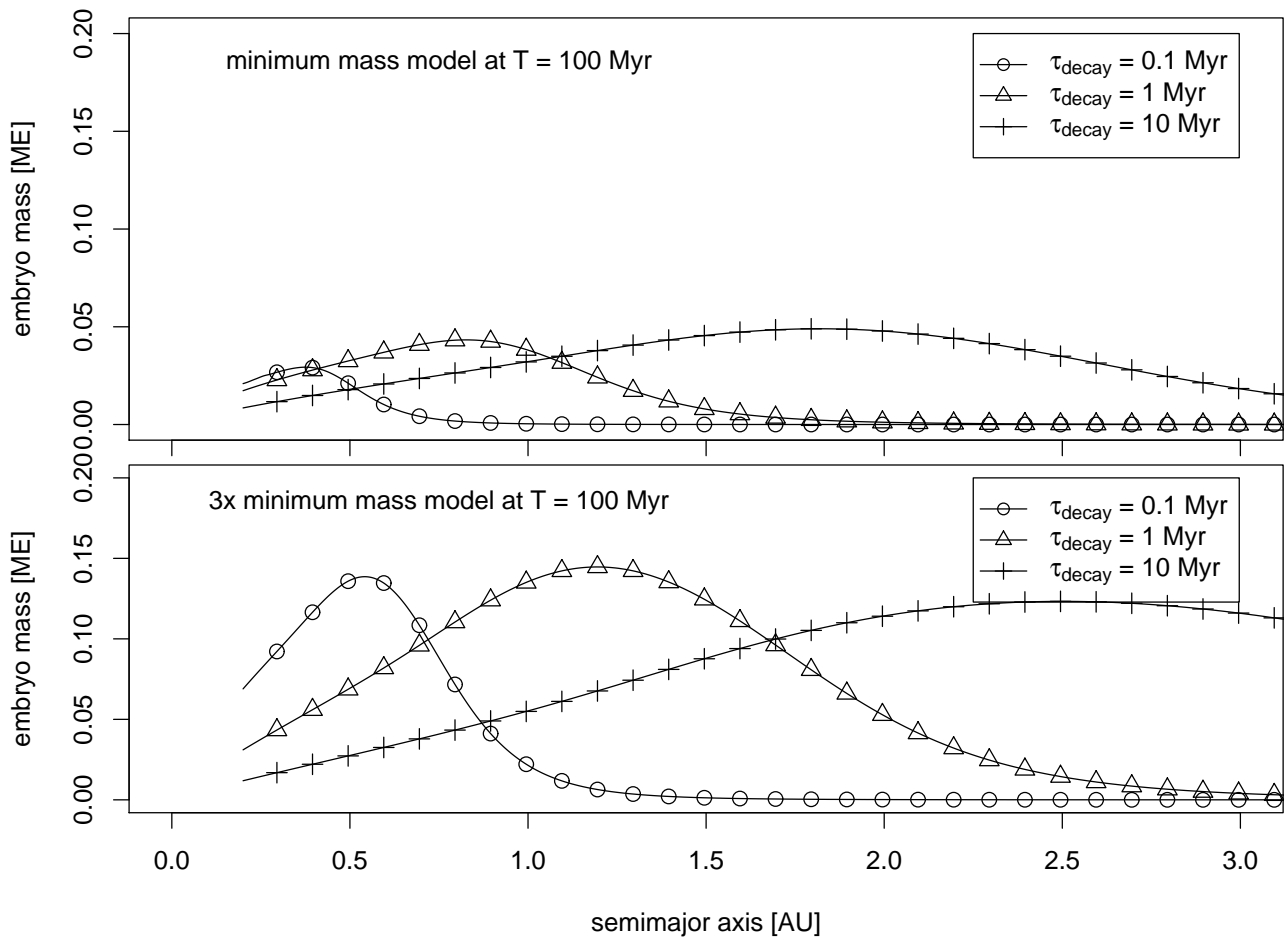}
\caption{Evolution of nominal and enhanced models with migration and dissipation.
\label{figure:ed_4}
}
\end{figure}

This suggests that in order to obtain the appropriate amount of
material in the terrestrial region with nominal migration, one could
(1) introduce a dissipating disc and (2) enhance the disc above the
minimum mass model.  Figure \ref{figure:ed_4} compares the end state
after 100 Myr of evolution for the minimum mass model with various
$\tau_{\rm{decay}}$ values with the same for a disc enhanced by a
factor of three.  The enhanced case both moves the point of maximum
embryo mass outwards and raises the maximum mass reached by almost a
factor of $\sim3$, but is otherwise qualitatively similar to the
minimum mass results.  From this and other runs (not shown) we
conclude that if we believe the nominal migration rates, then to
within the expected reliability of the model, disc enhancements of a
factor of several and dissipation times of $\sim$1 Myr should suffice
to keep enough prototerrestrial mass in the region.

Unfortunately these analytic models are sufficiently simplified that
they miss many interesting dynamical interactions and so cannot
predict the detailed behaviour of the embryos.  This is especially
true at the later stages when the embryo eccentricities are no longer
being damped by the gas.  The assumption that $b\!\sim\!10$ also
breaks down (see \S\ref{bevol}) as the embryo mass increases, and
therefore semi-analytic estimates of the mass in embryos remaining at
Myr timescales may be unreliable.  We therefore want to construct
N-body realizations of these models and integrate them directly, as we
discuss next.


\subsection{N-body modifications to Semi-analytic model}
\label{subsection:nbody_mod}
For consistency, we carry as much of the above approach as possible
through to our N-body code.  There are nevertheless modifications in
both the aerodynamic and the tidal drag formulae.

First, instead of the approximation for the aerodynamic drag in 
eq.\,(\ref{eq:v_aero}), we consider the planetesimal migration induced by
aerodynamic drag

\begin{equation}
\label{eq:v_m4nbod}
{\bf{a}}_{\rm{aero}} = \frac{d{{\bf{v}}}}{dt} = \frac{{\bf{v}} -
{\bf{v_{\rm{gas}}}}}{\tau_{\rm{aero}}}
,
\end{equation}

where $\bf{v}$ is the Cartesian velocity of the object,
$\bf{v_{\rm{gas}}}$ the velocity of the gas at the object's position,
and $\tau_{\rm{aero}}$ is given by eq.\,(\ref{eq:tdamp}).  As before, we
assume that $v_{\rm{gas}} = v_{\rm{kep}} \sqrt{1 - 2 \eta}$.

Second, for the tidal drag, we use the full approach of \cite{paplar},
which the authors developed to handle the case where a protoplanet's
eccentricity can be greater than the scale height-to-semimajor axis
ratio.  They derive timescales for semimajor axis damping $t_a$ and
for eccentricity damping $t_e$:

\begin{eqnarray}
\label{eq:ta}
  t_a = \frac{1}{c_a} \, 
  \left(\frac{a^3}{G M_\Sun}\right)^{1/2}
  \left(\frac{Z_1}{\rm{AU}} \, {\left(\frac{a}{\rm{AU}}\right)}^{1/4} \right)^{2} 
  \left(\frac{\Sigma_{\rm{gas}} \pi a^2}{M_\Sun}\right)^{-1} \,
  \nonumber\\
  \left(\frac{M}{M_\Sun}\right)^{-1} \,
  \left(
  \frac
      {1 + (e / 1.3) \, (Z_1/{\rm AU})^{-5} \, (a/\rm{AU})^{-5/4}}
      {1 - (e / 1.1) \, (Z_1/{\rm AU})^{-4} \, (a/\rm{AU})^{-1}}
      \right)
,
\end{eqnarray}

\begin{eqnarray}
\label{eq:te}
  t_e = \frac{1}{c_e} \, 
  \left(\frac{a^3}{G M_\Sun}\right)^{1/2}
  \left(\frac{Z_1}{{\rm{AU}}} \, {\left(\frac{a}{\rm{AU}}\right)}^{1/4} \right)^{4} 
  \left(\frac{\Sigma_{\rm{gas}} \pi a^2}{M_\Sun}\right)^{-1} \,
  \nonumber\\
  \left(\frac{M}{M_\Sun}\right)^{-1} \,
  \left(1 + \frac{e}{4} \,
  \left(\frac{Z_1}{\rm{AU}}\right)^{-3} \, 
  \left(\frac{a}{\rm{AU}}\right)^{-3/4}
   \right)
,
\end{eqnarray}
where (as before) $c_a$ is the migration efficiency, and $c_e$ is the
analogous damping efficiency.

They also argue that if the inclination damping timescale is not
significantly shorter than the eccentricity damping timescale then it
plays little role in the equilibrium state; we set $t_i = t_e$ for
simplicity.  From these we can find the acceleration on an object due
to tidal damping of semimajor axis and random velocity, namely
\begin{equation}
\label{eq:a}
  {\bf{a}}_{\rm{tidal}} = -\frac{\bf{v}}{t_a} - \frac{2
  ({\bf{v}} \! \cdot \! \bf{r})}{r^2 \, t_e}
  - \frac{2 ({\bf{v}} \! \cdot \! {\bf{k}})
  {\bf{k}}}{t_i}
,
\end{equation}
where $\bf{r}$, $\bf{v}$, and $\bf{a}$ are Cartesian position,
velocity, and acceleration vectors, respectively (with $r$ as the
magnitude of the radial vector) and $\bf{k}$ is the unit vector in the
vertical direction.

In the code (to be discussed in \S\ref{section:nbody_methods}), it is
equations (\ref{eq:v_m4nbod}) through (\ref{eq:a}) which are used to
incorporate the interaction between the embryos and planetesimals and
the gas disc.


\section{N-body Methods}
\label{section:nbody_methods}

We perform the numerical integrations using a parallel implementation
of SyMBA \citep{dll} called miranda.  SyMBA is a mixed-variable
symplectic integrator based on the N-body map of \cite{wishol} (see
also \citealt*{kyn}) which has been improved to accurately handle
close encounters between particles.  As is done in the analytic model,
we consider two classes of object, embryos (which gravitate and can
accrete planetesimals) and field planetesimals (where we neglect all
self-interactions.)  We have made several modifications to the
original algorithm, mainly to include the effects of the gas disc.  As
introduced in \S\ref{subsection:nbody_mod}, both embryos and
planetesimals suffer the aerodynamic drag of eq.\,(\ref{eq:v_m4nbod})
and the embryos further suffer semimajor axis decay and random
velocity damping according to the prescriptions of (\ref{eq:ta}),
(\ref{eq:te}), and (\ref{eq:a}).  Accelerations resulting from the gas
disc are treated as operators surrounding the (possibly recursively
subdivided) drift operator in the basic democratic heliocentric step
\citep{dll} after the manner of \citet{edphd}.  However, in practice
they are not applied during close encounters as the model for the gas
density is almost certainly inapplicable in such situations, and we
want to prevent artificial formation of tight binaries.  Note that we
neglect the gravitational potential due to the gas disc.

One important approximation made in our N-body treatment of the field
is the use of `super-planetesimals'.  That is, in representing the
field by a discrete number of particles, we use a large mass
$m_{\rm{sp}}$ for each object and imagine it represents an aggregate
of underlying planetesimals of mass $m$.  The aerodynamic drag that
the object feels is that which the underlying $m$-mass object would,
but when a merger occurs the mass $m_{\rm{sp}}$ is used.  Thus the
integrated field objects are serving as dynamical tracers.
Theoretical considerations \citep*[e.g.][]{bt} show that the accretion
rate of an embryo embedded in a planetesimal field should depend only
the product of the number density of the planetesimals and their mass,
provided that the collision timescale is much shorter than the embryo
growth timescale.  Numerical experiments varying the ratio 
$m_{\rm{sp}}/m$ over orders of magnitude confirm that the accretion is 
insensitive to the exact value of $m_{\rm{sp}}$ in our regime given 
that the characteristic embryo mass $M \gg m_{\rm{sp}}$ and the number of 
super-planetesimals is large.

It remains true that as a consequence of this approximation, at early
times when the ratio of typical embryo mass $M$ to $m_{\rm{sp}}$ is at
its lowest, we are inaccurately treating the effects of dynamical
friction.  However, the embryos are on circular, non-inclined orbits
in any event due to the strong tidal eccentricity damping, and the
aerodynamic drag is damping the planetesimals' random velocities,
which suppresses the frictional stirring.  As the drag becomes less
effective due to the dissipation of the disc, the $M/m_{\rm{sp}}$
ratio quickly increases, thereby reducing the inaccuracy.

We adopt as starting conditions $M/m_{\rm{sp}} \geq 5$ (where $M$ is
the characteristic initial embryo mass).  Note that this differs from
the approximation often made by Kokubo and Ida where they instead
increase the accretion radius of their particles by a factor $f$,
thereby artificially speeding up their simulations by a factor
$f^{-\beta}$ for some $\beta \in [1,2]$ (c.f.~\citealt{ko1};
e.g.~\citealt{ko_div}).

We also assume perfect accretion, which avoids the problem of
collisional cascades (and the resulting increase in particle number,
catastrophic in a nonstatistical code) at the cost of inaccuracy in
the mass spectrum; effects of this and other simplifications are
discussed in \S\ref{caveats}.

\section{Simulations}
\label{section:sims}
\subsection{Generating initial conditions for the N-body simulations}

Since computer power is limited, we wish to begin our integrations as
late in the second stage as possible.  We will use the semi-analytic
model of \S\ref{section:model} to evolve the disc from our nominal T=0
start until the number of embryos (i.e. the predicted number of
embryos; recall that $M(a)$ is continuous) is computationally
tractable.  This naturally produces a self-consistent gradient in
embryo masses with the appropriate field population.  We then
discretize $M(a)$ and $\Sigma_m(a)$, yielding a set of embryos and a
large number of super-planetesimal field particles, typically
$\sim\!8500$ (subject to the constraints on mass discussed in the
previous section), which constitute our N-body realization of the
initial conditions for our simulations.

To be specific, we set embryo and planetesimal densities $\rho_M =
\rho_m = 3.0 \,\rm{g\, cm^{-3}}$, and take the underlying physical
planetesimal mass to be $m = 2.1 \!\cdot\! 10^{-6} M_\Earth$
(i.e. planetesimal radius $r_m$ = 100 km at 3.0 $\rm{g \,cm^{-3}}$).  We
set the efficiencies of aerodynamic drag and tidal eccentricity
damping to unity, $C_D = 1, c_e = 1$, and set $\eta = 0.001$ at 1 AU.
We take $M_0$, the initial embryo mass at $T = 0$ for the semi-analytic
model, to be $0.0015 M_\Earth$, which is much larger than $m$ but two
orders of magnitude below the likely final values $M \geq 0.1
M_\Earth$.  By introducing this larger $M_0$ we advance the evolution
(especially of the embryos at larger $a$) forward in time.  This
procedure is justifiable to the extent that the evolution we are
bypassing is merely local accretion prior to any significant embryonic
tidal migration.  Thus, embryos in a simulation starting with smaller
embryonic seeds would largely `catch up' in mass to those in our
simulations by the time significant migration sets in.

We evolve this model until $T = 0.3$ Myr, which produces a
satisfactory ($\sim40-50$) number of embryos, and is early enough that
the use of a fixed $b$ of 10 is acceptable.  In this early phase, the
runaway growth timescale (to return the system to oligarchy if the
spacing is too large) and the scattering timescale (to do the same if
the spacing is too small) are both short.  Relaxation to the empirical
equilibrium conditions begins almost immediately in the simulations,
and we do not believe that our results are particularly sensitive to
the fine details at the start.

From this $T = 0.3$ Myr frame, objects are constructed from $a_0$ =
0.75 AU to $a_1 \approx$ 2.5 AU with embryo masses derived from $M(a)$
(objects placed inwards, from 0.50 AU to 0.75 AU, quickly fall off the
inner edge of the simulation at 0.40 AU.)  This sets the maximum mass
of a field object $m_{\rm{sp}}$ (see \S\ref{section:nbody_methods})
and from $\Sigma_m(a)$ the field objects can be built.  Initial
eccentricities and inclinations for embryos are set to 0.001 and
0.0005, respectively.  Planetesimal eccentricities are drawn from a
Rayleigh distribution of scale $e_{\rm{eq}}$ using
eq.\,(\ref{eq:eccequil}), and we enforce $i = e/2$.  (We observe that
eq.\,(\ref{eq:eccequil}) consistently predicts an equilibrium
eccentricity which is slightly larger than the value to which the
simulations rapidly relax.  Experiments suggest that this is due to an
overprediction of the stirring, whether the viscous or friction term,
and not an underprediction of the effects of aerodynamic drag.  This
is consistent with the comments of \cite{si} on stirring efficiency.)
All angles are randomized.

As discussed above, the implications of the simplified models of
\S\ref{section:model} are that to preserve sufficient mass we will
need to increase the surface density, and therefore (keeping in mind
possible enhancements in the outer solar system) we focus on discs
$\sim\!3$ to 4 times the minimum mass disc.  Specifications of the
resulting conditions are summarized in table \ref{table:sim1}.  We use
values for $\tau_{\rm{decay}}$ of 0.5, 1.0, and 2.0 Myr, and migration
efficiency $c_a$ of 0.25, 0.50, and 1.00.  For concreteness we choose
the $\tau_{\rm{decay}}$ = 1 Myr, $c_a = 0.5$ case to discuss below.

It was not obvious to what absolute time $T$ the simulations would
need to be advanced: after some experiments we chose a value of $T =
20$ Myr, when the gas has effectively vanished in all our simulations
and the inter-embryo spacing is large enough that the timescale for
mutual interaction is long.  Note that we do not include Jupiter and
Saturn in this simulation.

For example, fig.\,\ref{figure:C2_1} shows the initial (i.e. $T$ = 0.3
Myr) conditions for the N-body simulation produced by the model for
run C2.  We see the clear decrease in embryo mass with increasing $a$
characteristic of early times.  In the innermost region, 0.70-0.80 AU,
roughly 50\% of the material is contained in embryos, and in the
outermost region (2.4-2.5 AU) this proportion drops to less than 5\%.
(We define $f_{\rm emb}$ as the fraction of total mass in a region
which is in the form of embryos.)

The majority of the simulations were run at the Canadian Institute for
Theoretical Astrophyics on the McKenzie parallel machine, a Beowulf
cluster of $\sim\!256$ dual-processor 2.4 GHz Linux boxes, and the
remainder were integrated on local machines at Queen's University.

\begin{table}
\label{table:sim1}

\begin{center}
\caption{Simulation Parameters}\label{simspecs}
\end{center}
\begin{tabular}{ccccccc}
\tableline
Name & $f_{\rm enh}$ & $p$ & $\tau_{\rm{decay}}$ [Myr] & $c_a$ & $N_M$ & $N_m$\\
\tableline
C1  & 3.0 & -1.0 & 1.0 & 0.25 & 47 & 7987\\
C2  & 3.0 & -1.0 & 1.0 & 0.50 & 47 & 7987\\
C3  & 3.0 & -1.0 & 1.0 & 1.00 & 47 & 7987\\
C4  & 3.0 & -1.0 & 2.0 & 0.25 & 47 & 7876\\
C5  & 3.0 & -1.0 & 2.0 & 0.50 & 47 & 7876\\
C6  & 3.0 & -1.0 & 2.0 & 1.00 & 47 & 7876\\
C7  & 3.0 & -1.0 & 0.5 & 0.25 & 48 & 8451\\
C8  & 3.0 & -1.0 & 0.5 & 0.50 & 48 & 8451\\
C9  & 3.0 & -1.0 & 0.5 & 1.00 & 48 & 8451\\
C10 & 4.0 & -1.5 & 1.0 & 0.25 & 45 & 9570\\
C11 & 4.0 & -1.5 & 1.0 & 0.50 & 45 & 9570\\
C12 & 4.0 & -1.5 & 1.0 & 1.00 & 45 & 9570\\
C13 & 4.0 & -1.5 & 2.0 & 0.25 & 44 & 9120\\
C14 & 4.0 & -1.5 & 2.0 & 0.50 & 44 & 9120\\
\tableline
\end{tabular}

\tablecomments{$f_{\rm enh}$ and $p$ are defined through $\Sigma_{\rm
solid} = (f_{\rm enh} \cdot 7.1 \rm{\,g \, cm^{-2}}) (a/{\rm AU})^p$
where $\Sigma_{\rm solid}$ is the original surface density in solids
at T=0.  $\tau_{\rm{decay}}$ is the e-folding time of the gas
dissipation, $c_a$ is the migration efficiency factor, and $N_M$ and
$N_m$ are the embryo and field particle counts.  Each field particle
is a super-planetesimal as described in \S\ref{section:nbody_methods}.
}


\end{table}

\pagebreak

\subsection{Results}

\begin{figure}
\epsscale{.80}
\plotone{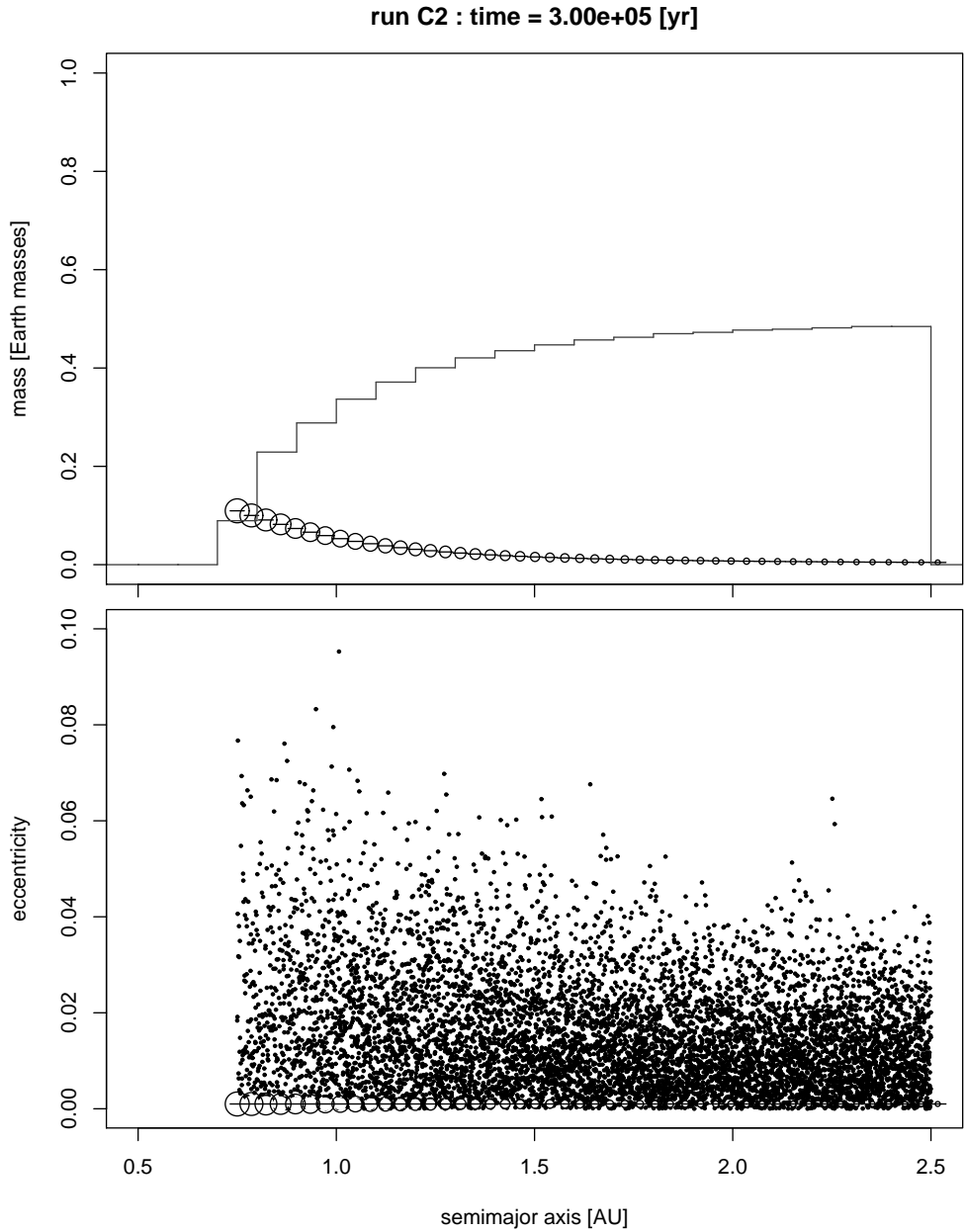}
\caption{Initial conditions at $T = 0.3$ Myr for model C2.  Large
circles correspond to embryos (of area proportional to mass) and small
circles to planetesimals; the lines through the large circles indicate
a width of 10 Hill radii. The broken lines on the top graph indicate
the amount of planetesimal material (in Earth masses) in a semimajor
axis bin of width 0.1 AU.
\label{figure:C2_1}
}

\end{figure}

\begin{figure}

\epsscale{.80}
\plotone{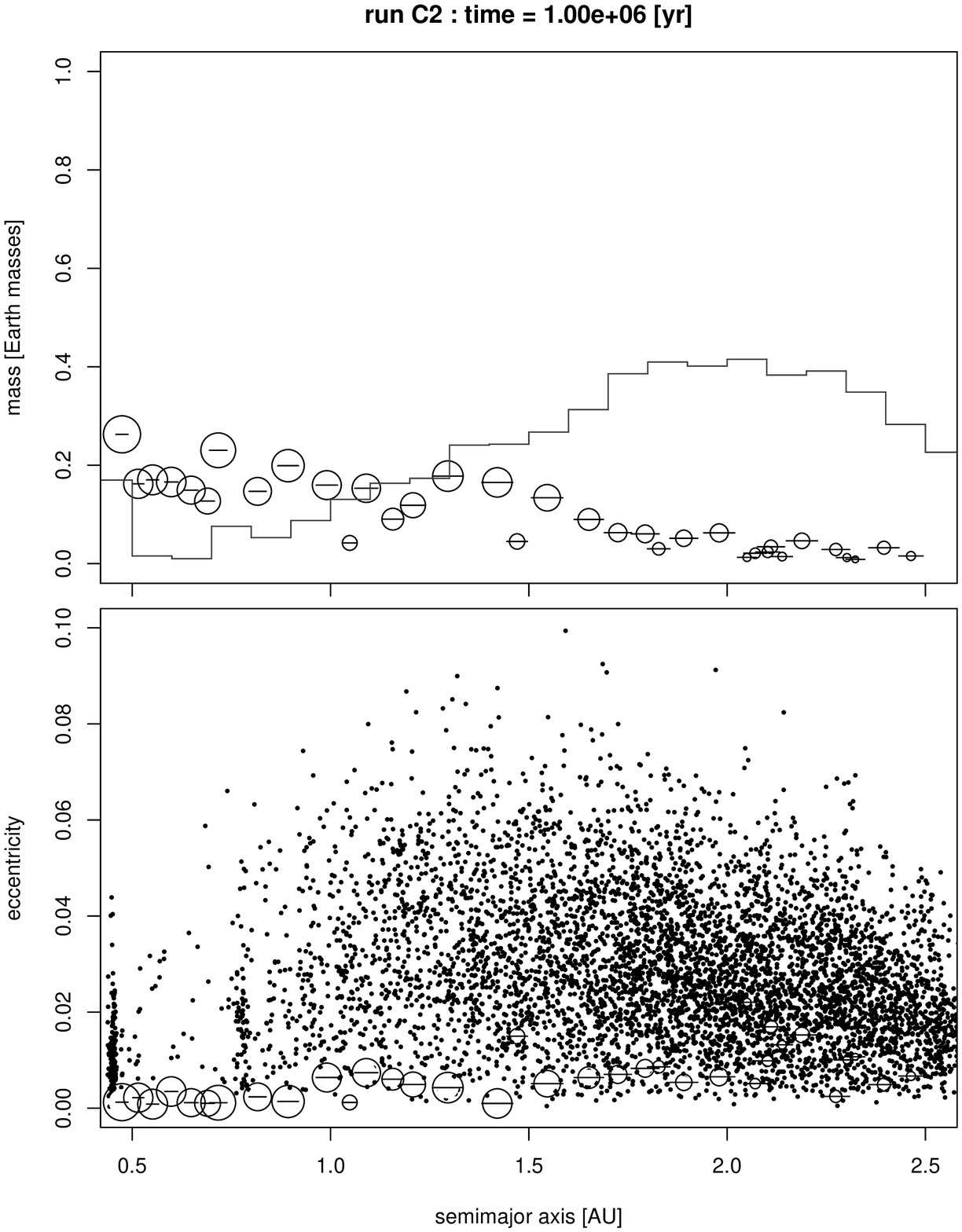}
\caption{Configuration at $T = 1$ Myr for model C2.
\label{figure:C2_2}
}

\end{figure}

\begin{figure}
\epsscale{.80}
\plotone{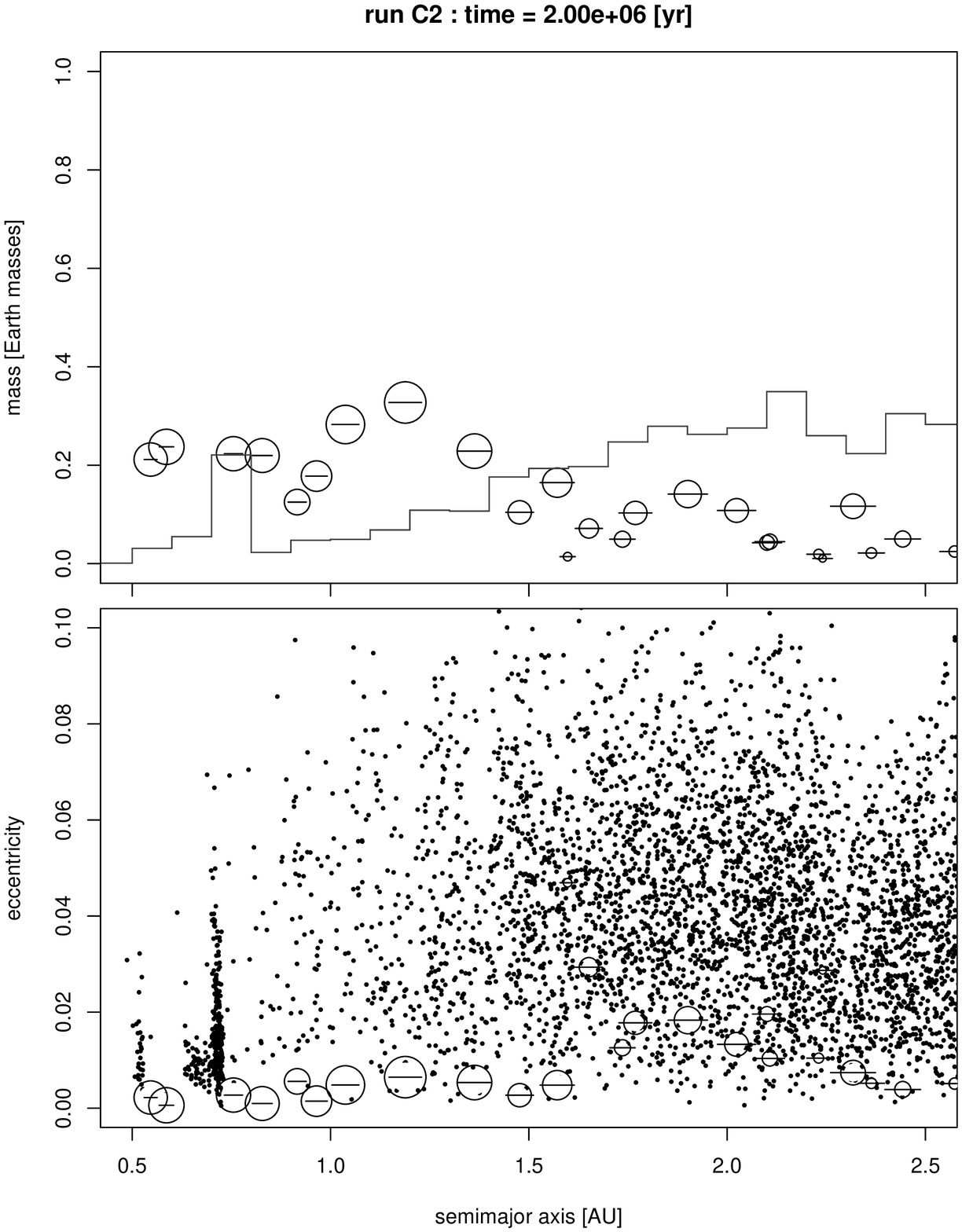}
\caption{Configuration at $T = 2$ Myr for model C2.
\label{figure:C2_3}
}

\end{figure}

\begin{figure}

\epsscale{.80}
\plotone{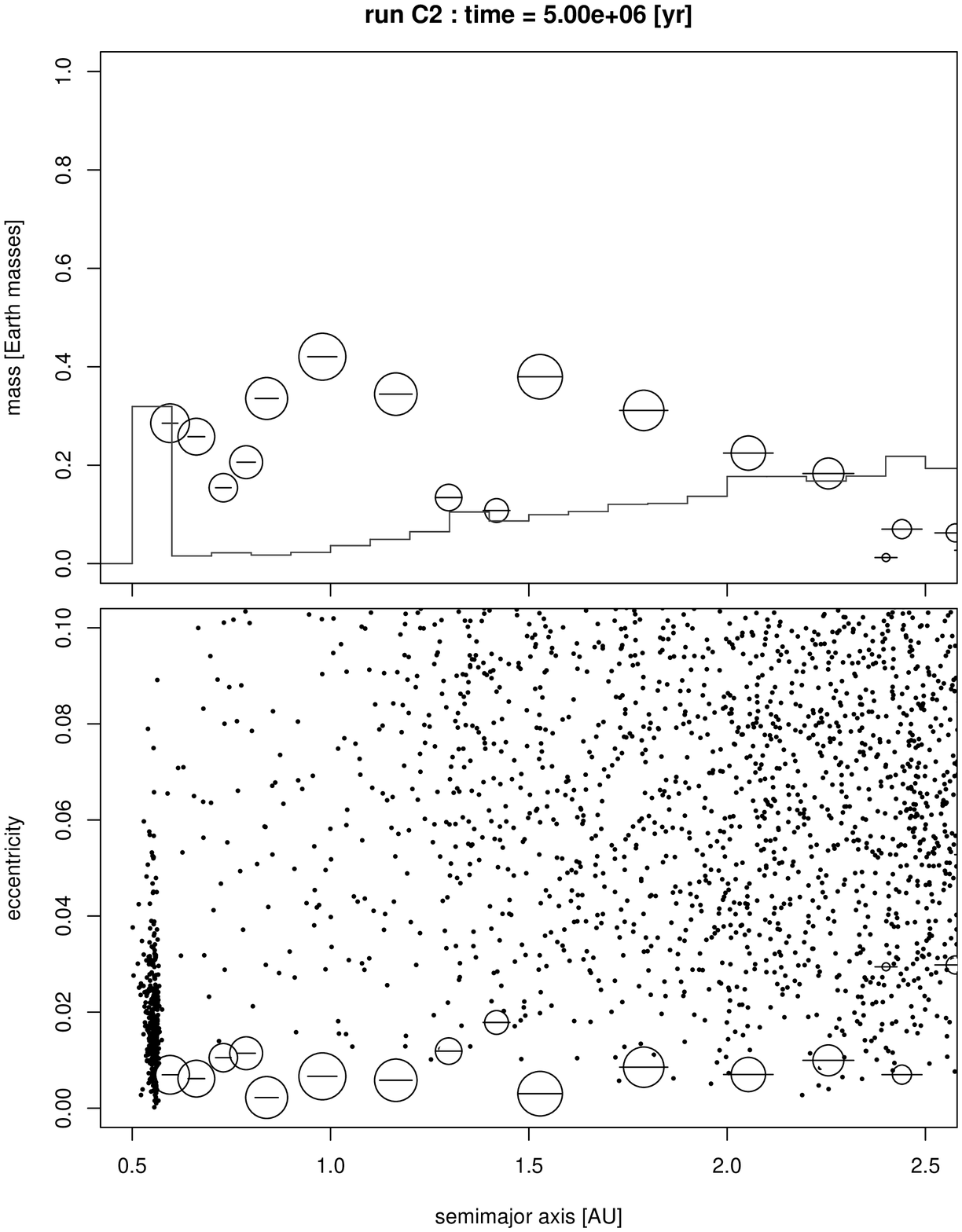}
\caption{Configuration at $T = 5$ Myr for model C2.\label{figure:C2_4}
}
\end{figure}

\begin{figure}

\epsscale{.80}
\plotone{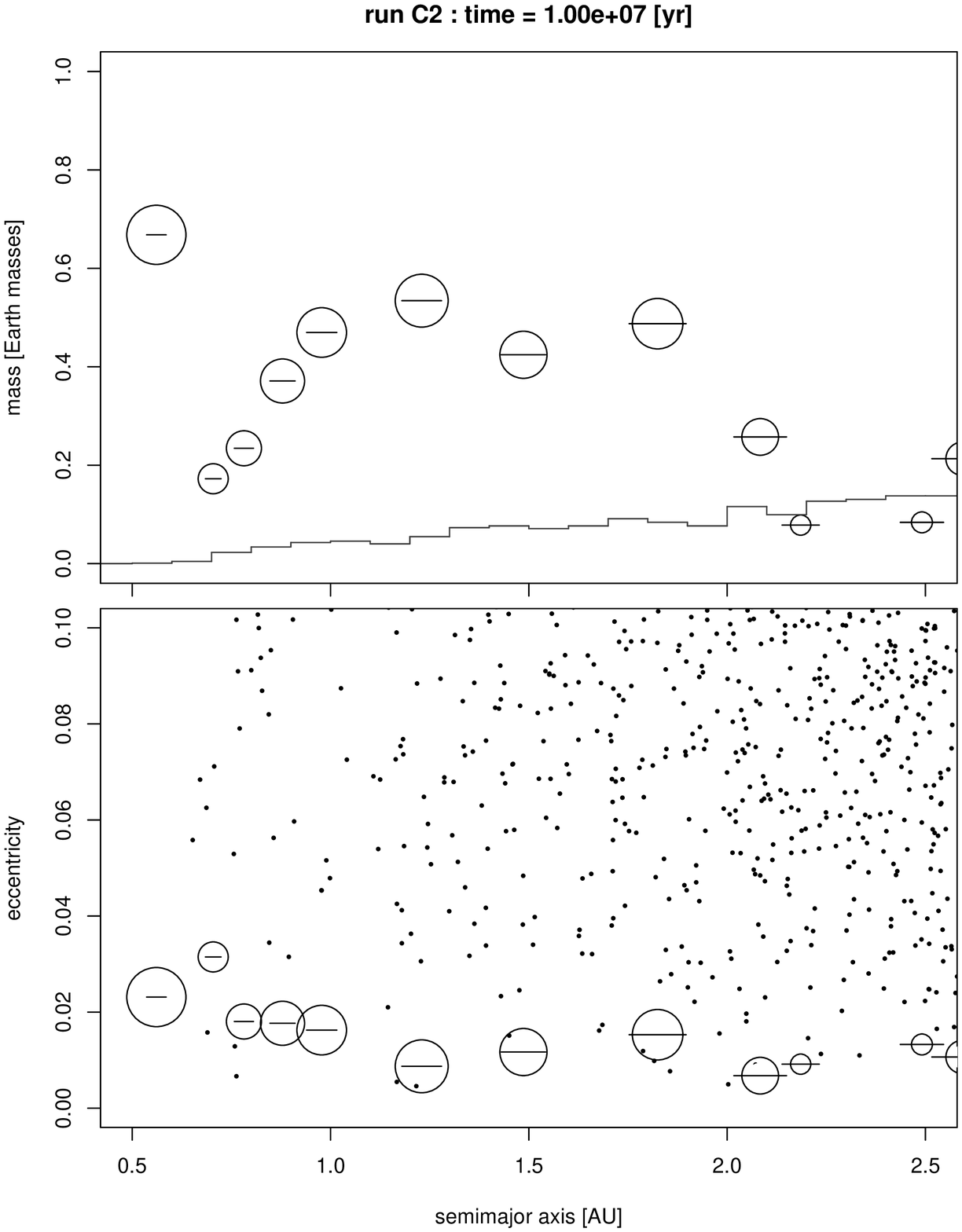}
\caption{Configuration at $T = 10$ Myr for model C2.\label{figure:C2_5}
}
\end{figure}

\begin{figure}

\epsscale{.80}
\plotone{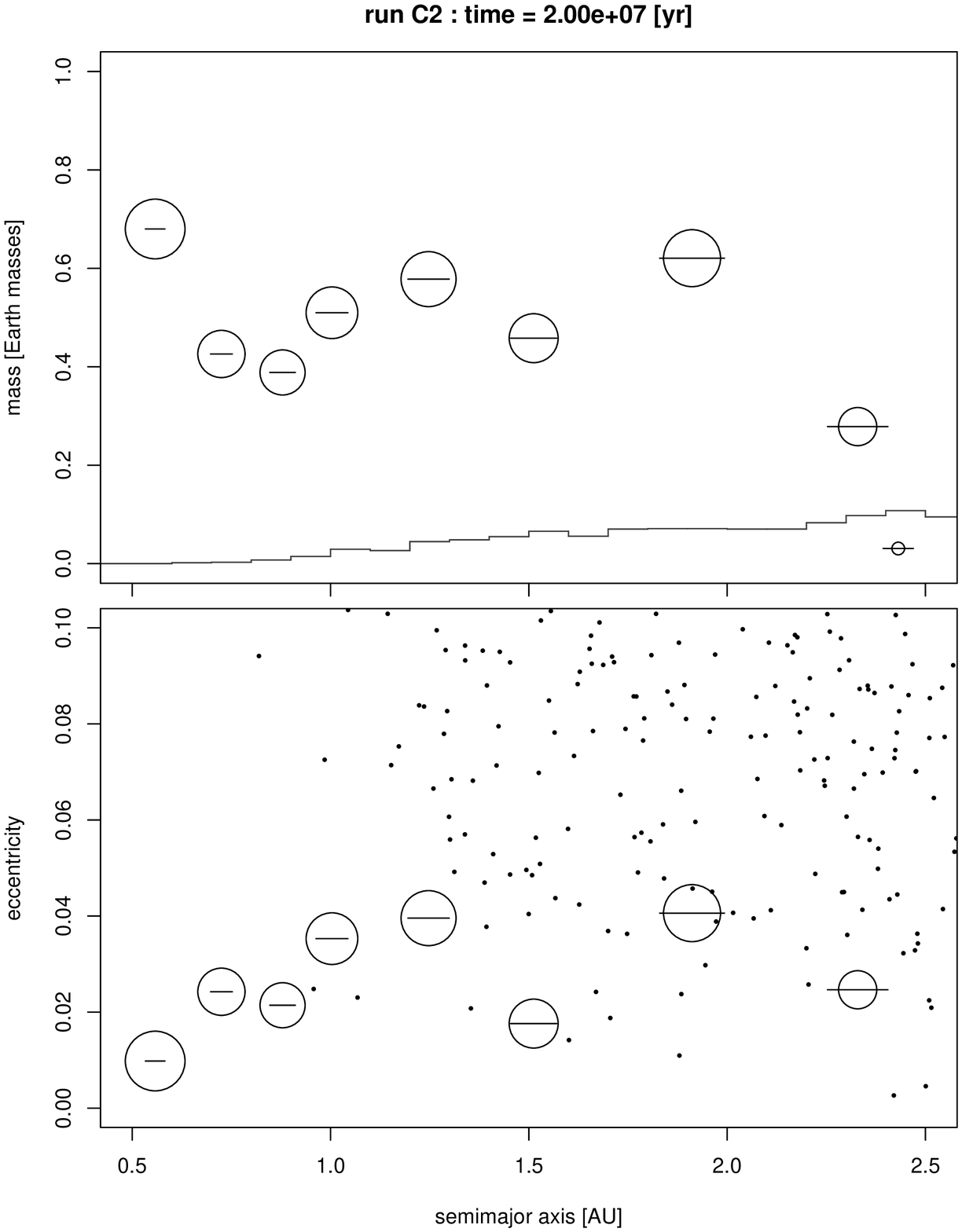}
\caption{Configuration at $T = 20$ Myr for model C2.\label{figure:C2_6}
}
\end{figure}

As an example, time slices in $(a,M)$ and $(a,e)$ from C2, a
representative run which demonstrated most of the typical behaviours,
are displayed in figures \ref{figure:C2_1} through \ref{figure:C2_6}.
In this simulation, $\Sigma_m = 21.3 \, \rm{g \,cm^{-2}} \, (a/\rm{AU})^{-1}$,
$c_a = 0.5$, and $\tau_{\rm{decay}}$ = 1 Myr.

After 0.7 Myr, in figure \ref{figure:C2_2}, at $T = 1$ Myr (1
e-folding time of the gas decay), the system has undergone
considerable evolution.  Many embryos have undergone mutual mergers
and some have migrated beyond the inner edge of the simulation and
been removed.  Figure \ref{figure:C2_2} shows wide ranges of roughly
oligarchic behaviour where neighbouring objects have comparable mass.
In the region from 0.5 to 1.6 AU, for example, most of the embryos
have mass $\sim0.18M_\Earth\pm0.02M_\Earth$.  At this time the region
where $f_{\rm emb} \simeq0.5$ extends from 1.0-1.5 AU.  The variation
of evolution timescale with semimajor axis is apparent, but both
embryo-field and embryo-embryo mergers have occurred beyond 1.5 AU.

In zones where an embryo has achieved some separation from its
neighbours, `Jacobi wings' can be formed (see \citealt{tanaka}).  Jacobi
wings are the characteristic configurations -- a curved V -- produced
in the $(a,e)$ plane by an embryo embedded in a disc of planetesimals
as it scatters them, so called because the shape of the scattering
paths of the field particles are determined by the the conservation of
the Jacobi integral in the restricted circular three-body problem.
One such wing is apparent interior to the embryo near 0.7 AU in
fig.\,\ref{figure:C2_3}, where the Jacobi wing has swept up a
considerable amount of mass ($0.2 M_\Earth$) during migration.

At this time, $0.43 M_\Earth$ of material has been removed from the
simulation for approaching the Sun too closely (0.4 AU being the
limit).  Ultimately $2.1 M_\Earth$ will be removed, and 90\% of this
is accomplished by 3 Myr.  The probable fate of the removed material
is discussed further in \S\ref{section:weiden}; there are good reasons
to believe that most of the material which escapes makes it to the
solar surface.  With migration, in order to leave enough mass in the
terrestrial region, it becomes necessary to make a substantial
sacrifice to the Sun.

At 2 Myr, in figure \ref{figure:C2_3}, when the gas has dropped to
$\sim\!1/7.4$ of its original density, more objects have been lost to
the inner edge, and those that remain from 0.5 to 1.5 AU retain
comparable mass (i.e. the deviation from the mean is less than a
factor of two) and have grown to a mean mass just over $0.2 M_\Earth$.
By 5 Myr (figure \ref{figure:C2_4}) the $f_{\rm{emb}} \lesssim 0.5$
region is now beyond 2.0 AU and we are approaching the end of the
transition phase and the beginning of our end stage (where
embryo-planetesimal accretion no longer plays a significant role.)
The mean embryo mass in the 0.5-1.5 AU region is $\sim\!0.25
M_\Earth$, brought down somewhat by the objects from 1.3 to 1.4 AU.
At this particular time the embryo spacing from 0.5 to 0.8 AU is
anomalously small, and this tightly-packed configuration quickly
evolves to a wider one.  At this time, after 5 e-foldings, the gas
density has dropped to $\sim\!1/150$ of its original value and gas
drag (whether aerodynamic or tidal) is not important.  The
corresponding increase in eccentricity pushes the accretion timescale
up substantially, and the dominant growth mechanism will now be
embryo-embryo mergers.

Between 10 Myr and 20 Myr (figures \ref{figure:C2_5} and
\ref{figure:C2_6}) we arrive at a long-lived configuration of embryos:
only one embryo merger occurs between 0.5 AU and 1.5 AU, and it occurs
before 14 Myr.  Evolution is continuing beyond 1.75 AU, and although
limited, the remaining planetesimals are gradually being consumed.
The mean embryo mass in the region of interest (0.5-1.5 AU) is
$\sim\!0.45 M_\Earth$, and the spacing varies from $b\sim\!15$ to
$b\sim\!25$, which is very well-separated.  The eccentricities of the
bodies ($\sim0.02-0.04$), which have been increasing over time, are now
many times above their tidally-suppressed values.  The region contains
$2.58 M_\Earth$ in the form of embryos and a residual planetesimal
field of $0.23 M_\Earth$ for a total of $2.81 M_\Earth$ in the zone
from 0.5-1.5 AU.

As previously noted, roughly $2 M_\Earth$ of material escapes the
simulation to the inside.  A substantial amount of mass remains beyond
the zone of interest, however: $3.55 M_\Earth$ of material is located
beyond 1.5 AU.  This is problematic, as there is nothing approaching a
$4 M_\Earth$ object between the Earth and Mars; we return to this
problem in \S\ref{transport}.


\begin{figure}
\epsscale{.80}
\plotone{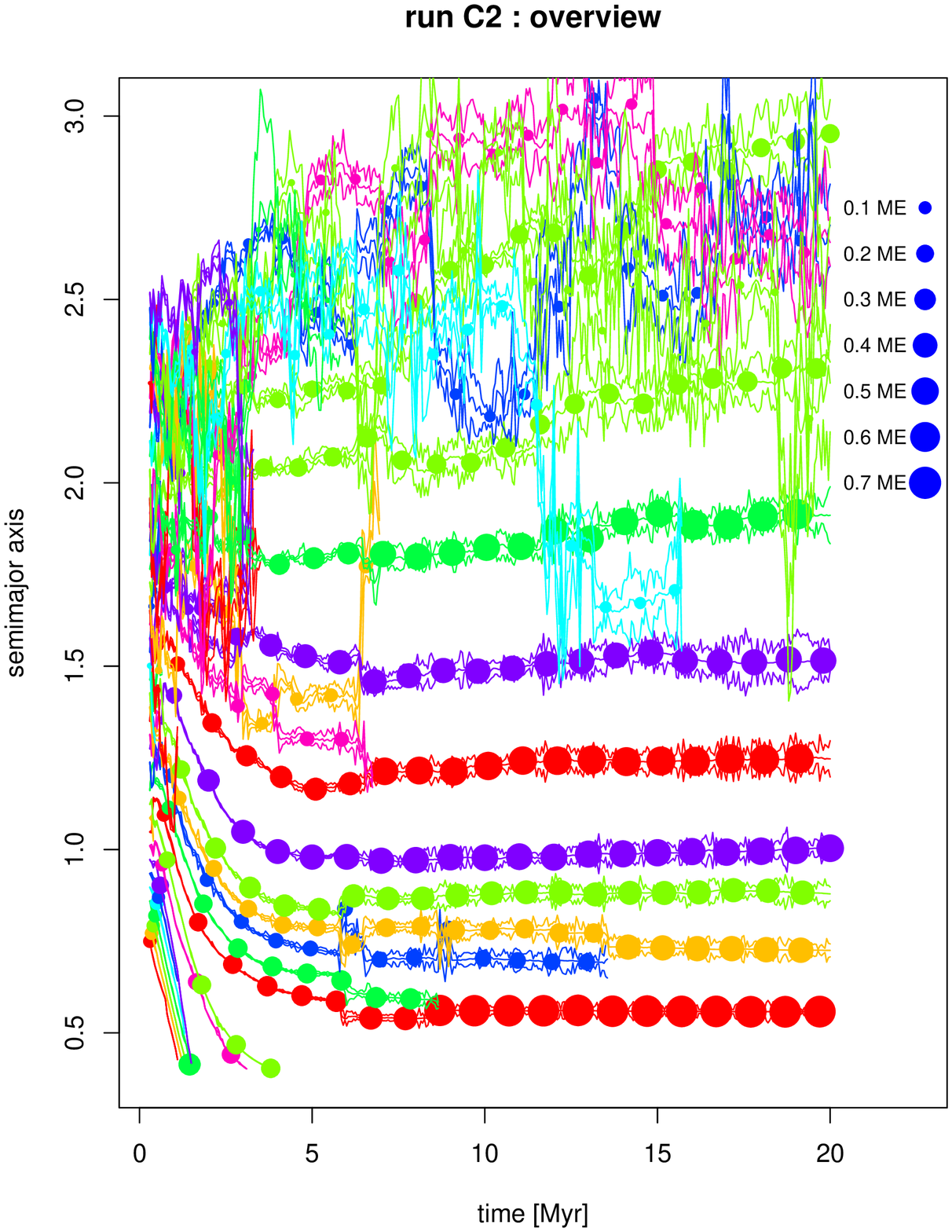}
\caption{Evolution of embryos with time in C2.  The area of the
circles scales linearly with embryo mass as indicated in the legend.
For each embryo three lines are drawn, corresponding to the osculating
values of the perihelion, semimajor axis, and aphelion.\label{figure:Mat1}}

\end{figure}

\begin{figure}
\epsscale{.80}
\plotone{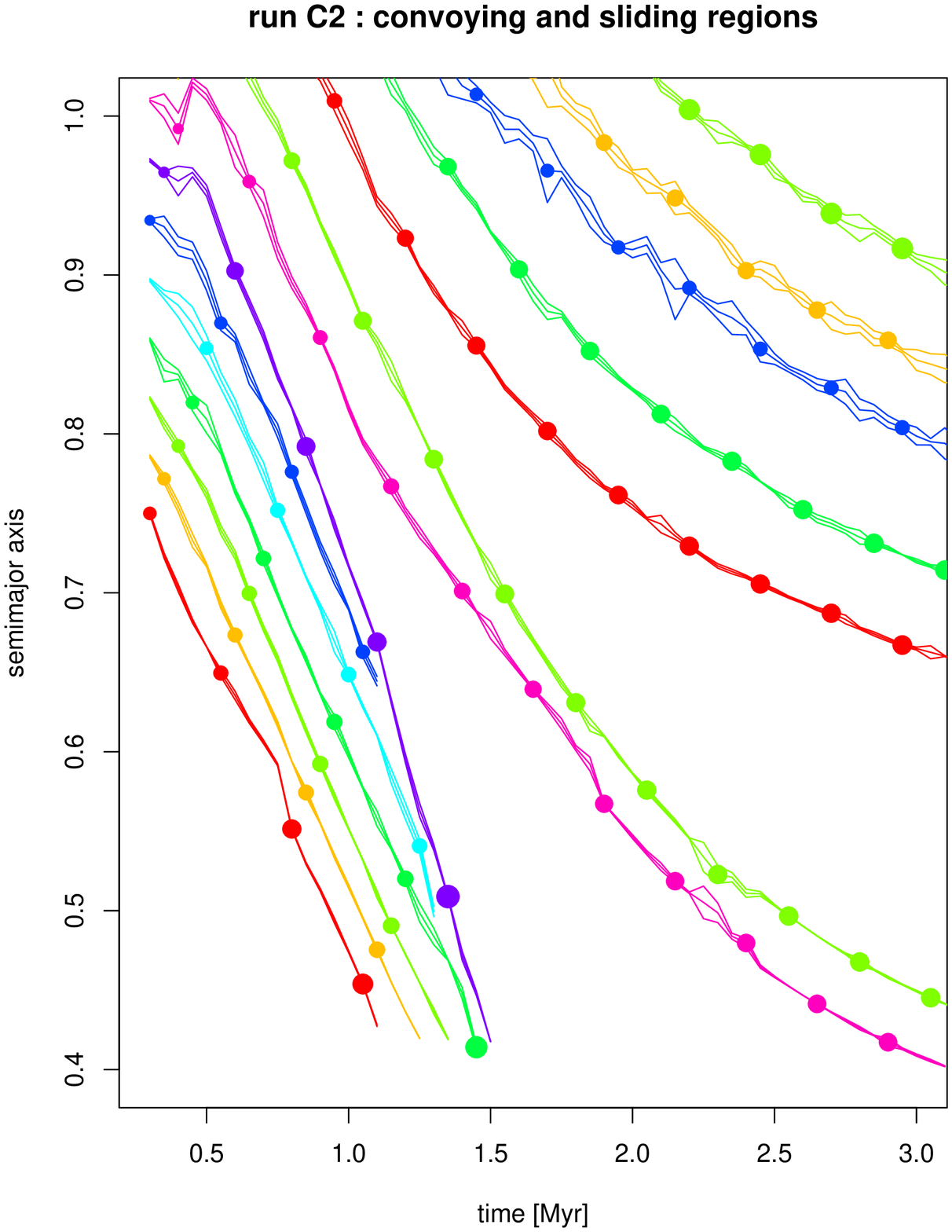}
\caption{The interior regions of C2 at early times. \label{figure:Mat2}}

\end{figure}

Viewing the embryo evolution in $(M,a,t)$ space is instructive, as
shown in fig.\,\ref{figure:Mat1}.  Three different regimes are
apparent: for $a <$ 1.0 AU the embryos head directly for the Sun
without scattering; for 1.0 AU $< a <$ 1.5 AU the embryos `slide'
(i.e.~collectively move in an orderly fashion, without orbit crossing)
toward their final destinations, slowing as the gas decay lowers the
migration rate; and for $a > 1.5$ AU the evolution is highly chaotic.
Figure \ref{figure:Mat2} magnifies the innermost regions.

At early times, when the embryo masses are low, the embryos can
reorder themselves substantially and move several tenths of an AU
without suffering a merger with another embryo.  However, as time
passes and the masses of the embryos grow (whether through accretion
of planetesimals or direct embryo mergers), the migration pulls
objects inwards at different rates.  This leads to embryo
segregration, and in some cases (typically where migration rates are
high) to the formation of groups of two to four members which we call
tidal convoys.  (Although not mentioned explicitly, this effect also
appears to have been present in the simulations of \citealt*{paplar};
see figure 6 in their paper.)

At any time in these convoys the objects are typically ordered such
that the mass increases outwards.  This is self-selection: if an inner
object is more massive than an outer one, differential migration will
cause the objects to diverge, breaking the convoy.  As one would
expect, the migration rate of the convoys is above that of the less
massive objects and below that of the most massive object as the outer
objects `push' on the inner ones.  In many of our simulations, as in
C2, these groupings are only apparent on the inner edges (where the
object masses and gas density are highest) as the decay of the gas
disc and corresponding weakening of the migration makes the convoys
harder to form and maintain.  If we set $\tau_{\rm decay} = \infty$,
tidal convoying becomes nearly inevitable, which raises serious
questions about the applicability of analytic oligarchy models to the
middle-to-late stage transition unless the gas is presumed to have
vanished.

In our simulations, while the gas remains, type I migration is
remarkably efficient at capturing embryos into first-order mean motion
resonances with each other, and the convoys usually show resonant
relationships (two-object pairs are especially susceptible.)  This
enhancement of resonance capture probability due to migration has been
discussed in a different context by previous authors (see
\citealt*{peale} for a possible explanation for the Laplace resonance
among the Galilean satellites.)  Most two-member tidal convoys are
observed to be stable, but arguments to that end
\citep*[e.g.][]{gladstable} are not easily generalized to the
multiple-planet case, and in practice convoys involving three or more
objects of different masses are eventually observed to merge.

Beyond the convoying region, fig.\,\ref{figure:Mat2} shows part of the
`sliding' region (and, conveniently, a pair showing intermediate
behaviour.)  As is apparent in fig.\,\ref{figure:Mat1}, when the masses
and gas densities are such that the spacings will stay roughly
constant during the migration, a group of embryos can move smoothly
from further out in the disc suffering only a few mutual mergers.
(Admittedly, in C2 the surface density of the gas disc varied as
$\Sigma \propto a^{-1}$, and one can show (see section
\ref{section:weiden}) that $da/dt$ is constant for a fixed mass in
this case.  However, in the sliding region the masses vary by a factor
of several, the gas density was decaying, and similar behaviours are
seen for the $\Sigma \propto a^{-3/2}$ case.  The effect is not a
peculiarity of this particular density law, although it is a
consequence of the near-constant migration rates that it and similar
laws provide.)

Finally, in the lowest-mass region, chaotic behaviour persists for the
entire length of the simulation, although it does begin to settle down
by the end.  In one case, the embryo excursion reached 1.3 AU.  Even
in this chaotic regime, the broad outlines of the oligarchic model
(roughly equally spaced embryos of comparable mass) were still
respected, despite the fact that the simplest oligarchic picture is
inapplicable as the embryos are being constantly reordered.  This may
imply that other mechanisms underly the oligarchic model's impressive
robustness \citep{gold}.

These three regimes (convoying, sliding, and chaotic) were observed
throughout the simulations -- although their locations and strengths
varied with mass, gas density, and stochastically-set spacing, as
expected -- and are likely frequent.

\subsection{Final Characteristics}
\label{subsection:final}

\begin{figure}

\epsscale{.80}
\plotone{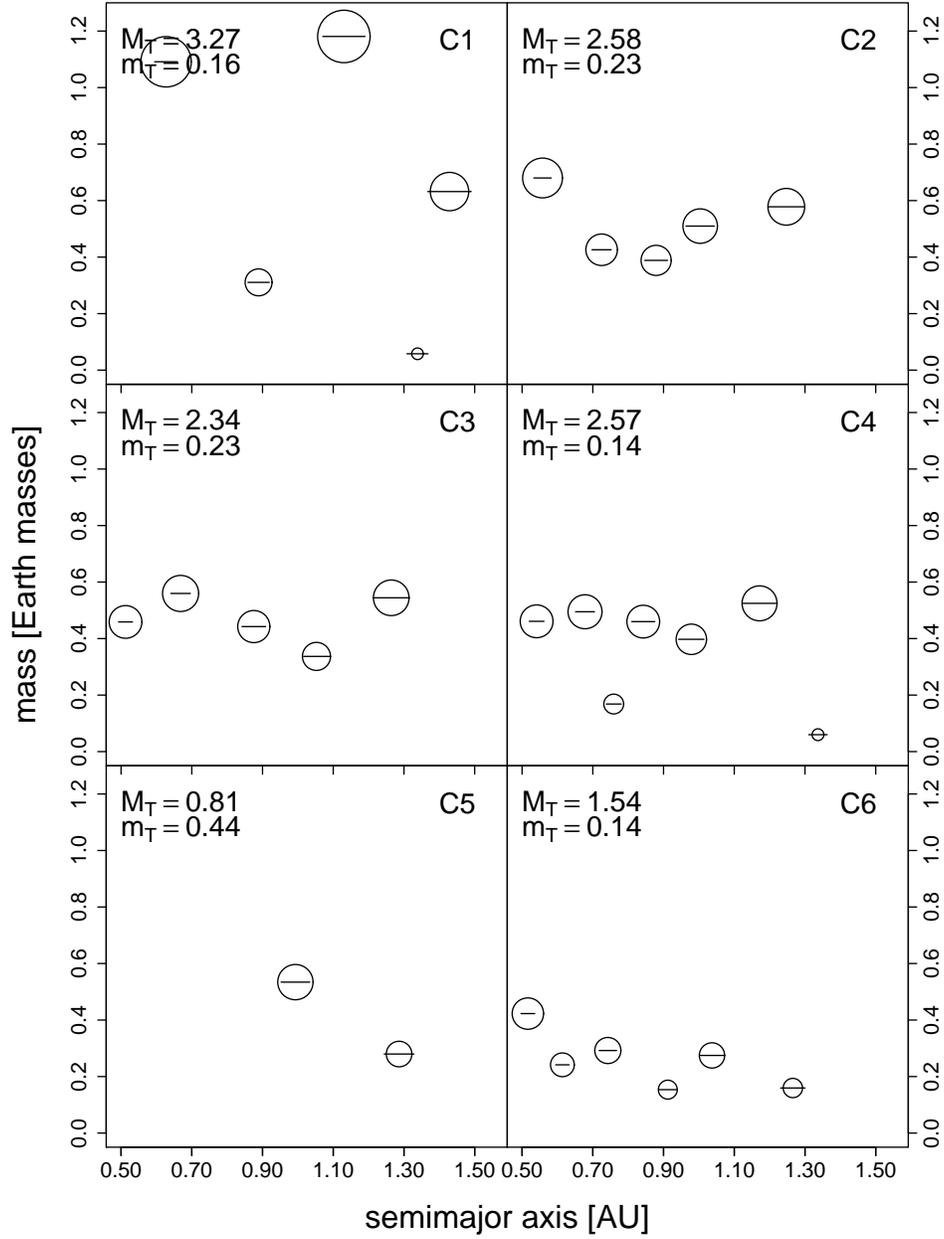}
\caption{
\label{figure:Cset1}
Final configurations in mass and semimajor axis for runs C1-C6.  $M_T$
and $m_T$ indicate (in Earth masses) the total mass in embryos and in
planetesimals, respectively, in the region [0.5 AU, 1.5 AU].}
\end{figure}

\begin{figure}

\epsscale{.80}
\plotone{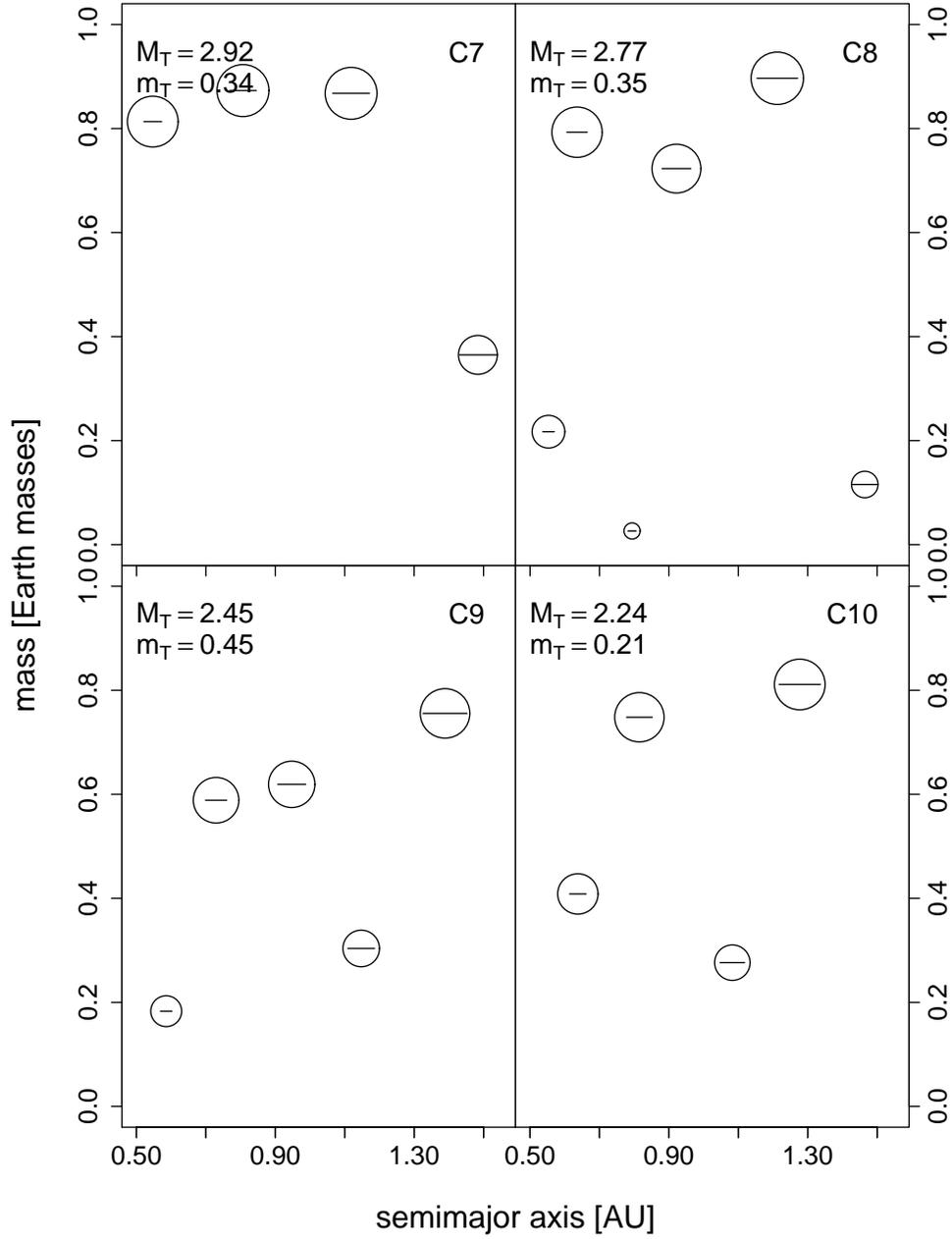}
\caption{
\label{figure:Cset2}
Same as fig.\ref{figure:Cset1}, but for runs C7-C10.}
\end{figure}

\begin{figure}

\epsscale{.80}
\plotone{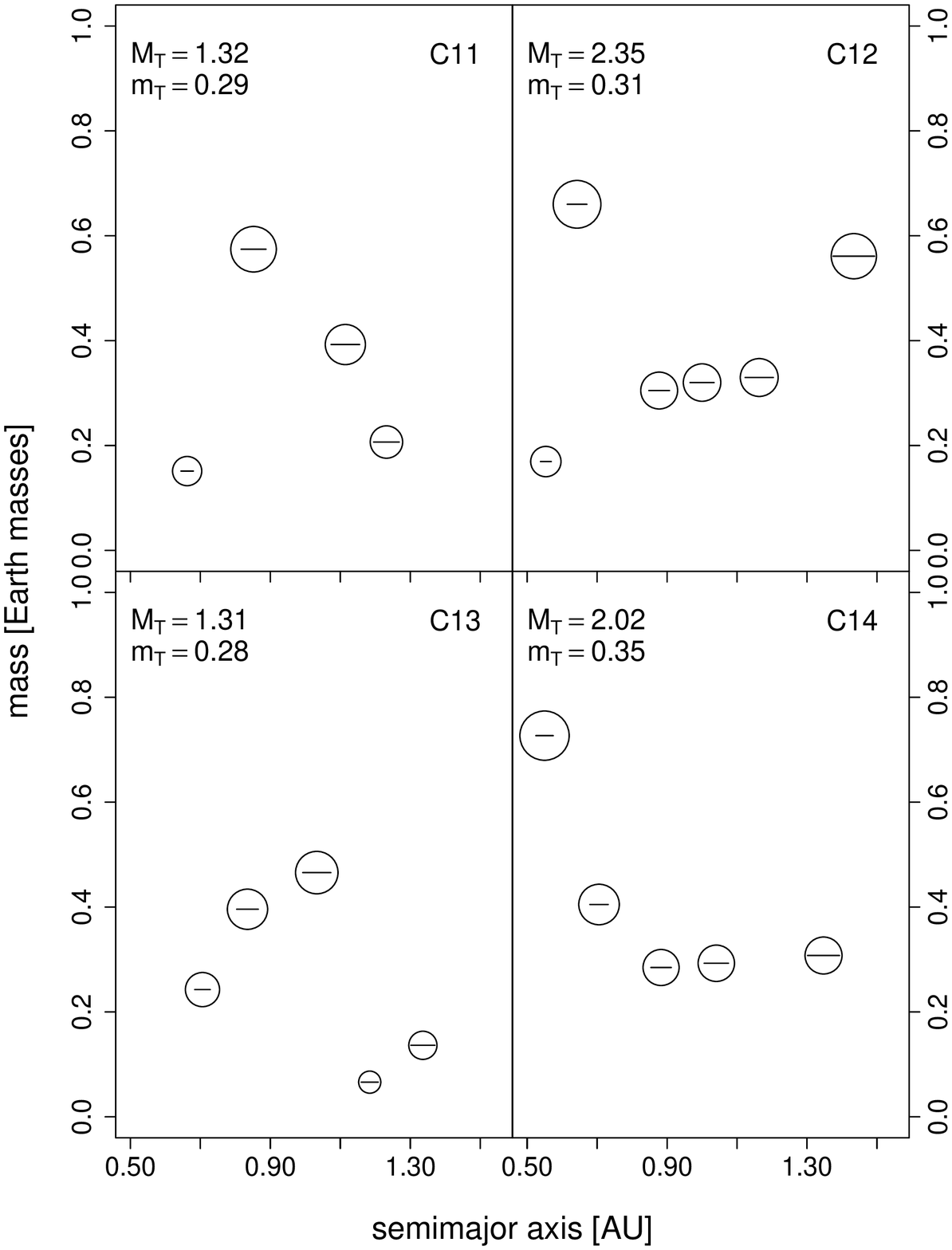}
\caption{\label{figure:Cset3}
Same as fig.\ref{figure:Cset1}, but for runs C11-C14.}
\end{figure}

Time slices of the final configurations for simulations C1-C14 are
plotted in figures \ref{figure:Cset1}, \ref{figure:Cset2}, and
\ref{figure:Cset3}.  

The resulting mass in the 0.5-1.5 AU region in the form of embryos
ranges from $0.81 M_\Earth$ to $3.27 M_\Earth$, and for planetesimals
from 0.14$M_\Earth$ to 0.44$M_\Earth$.  Some of the runs (C2, C3, C4,
C6) yield very oligarchic-looking outcomes where the variation in mass
between embryos is low, and the characteristic mass varies from 0.2 to
0.4 $M_\Earth$.  Others (C1, C7, C8, C9, C10) produce systems with
embryos of mass approaching $1 M_\Earth$, at least some of which are
candidate planets.  The remaining simulations are difficult to
classify, although most show regions oligarchic in appearance with a
few outliers (e.g.~C12, C14.)  C5 produced the least mass (embryo mass
$0.81 M\Earth$, field mass $0.44 M_\Earth$) for reasons to be
discussed in \S\ref{walls}, and C1 the most ($3.27 M_\Earth, 0.16
M_\Earth$).  By accident, C1 produced two dominant objects of
terrestrial mass -- Venus and Earth analogues.

\begin{figure}
\epsscale{.80}
\plotone{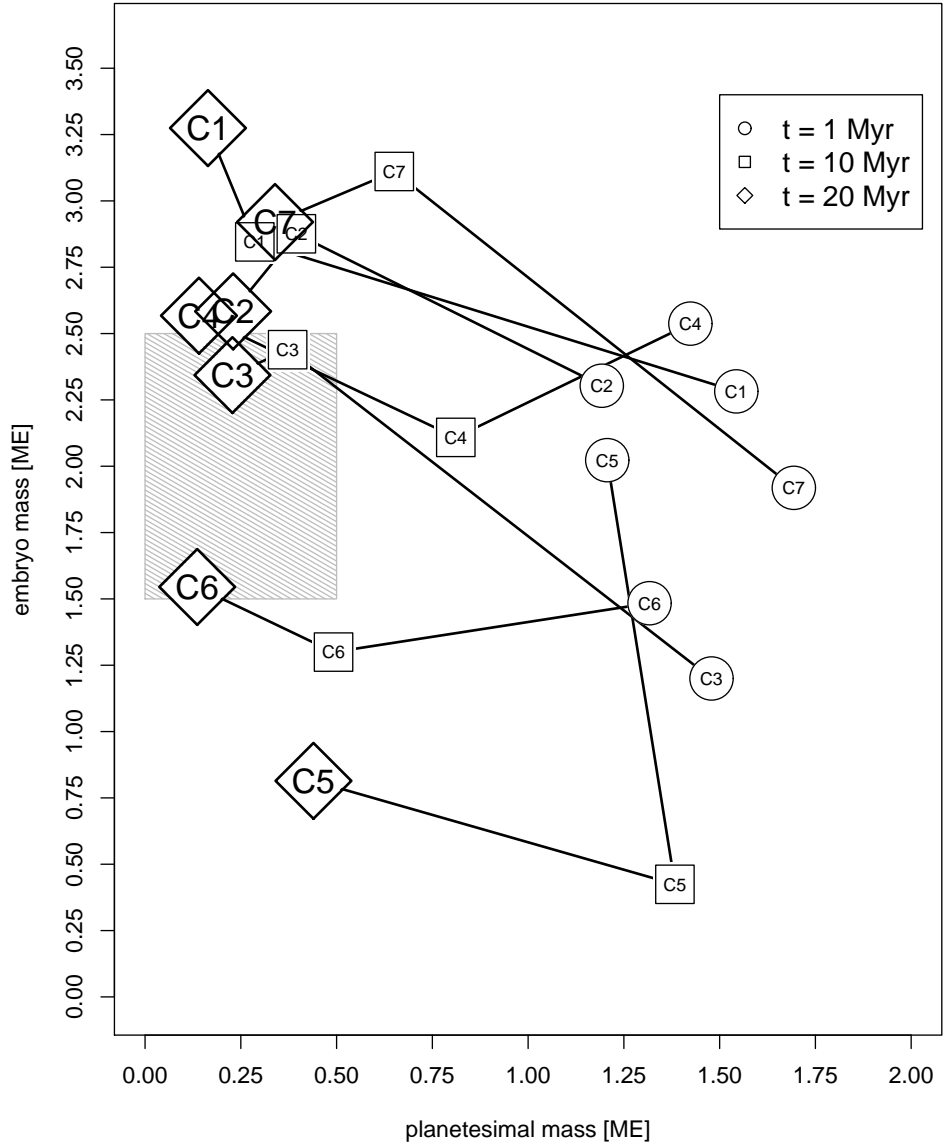}
\caption{
\label{figure:Mvsm1}
Embryo mass versus planetesimal mass within the 0.50-1.50
region at 1, 10, and 20 Myr for runs C1-C7.  The nominal target zone
(from 1.5 to 2.5 $M_\Earth$ in embryo mass and $\leq0.5 M_\Earth$ in
planetesimal mass) is shaded.}
\end{figure}

\begin{figure}

\epsscale{.80}
\plotone{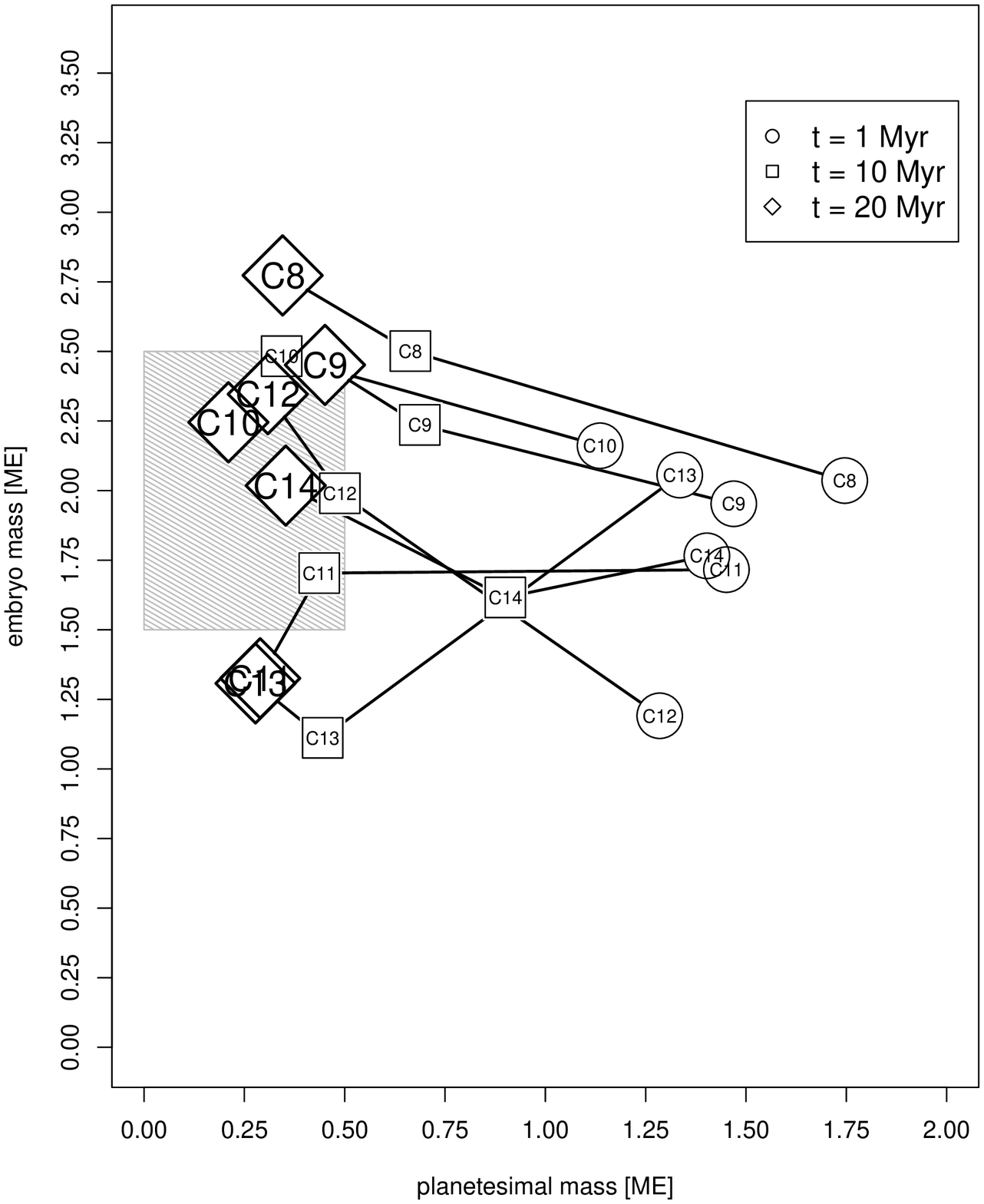}
\caption{
\label{figure:Mvsm2}
Same as fig.\ref{figure:Mvsm1}, but for runs C8-C14.}
\end{figure}

Figures \ref{figure:Mvsm1} and \ref{figure:Mvsm2} show the resulting
amounts of embryo and field mass for all simulations at 1, 10, and 20
Myr.  Most simulations show an increase in embryo mass with time, and
a substantial number of simulations landed in our nominal target
region of 1.5 to $2.5M_\Earth$ in embryo mass and $\leq0.5 M_\Earth$
in planetesimal mass.

The original expectation was for all the simulations to produce
oligarchic-looking results (as in C3), not direct planet production
(as in C1), but it is not surprising that with more massive embryos at
least some embryo-embryo mergers would occur as the gas vanished, and
it takes only a few such mergers at late times to produce a
substantial planet.  Of the fourteen simulations, four produced less
than two Earth masses (our nominal target) of material in the region
and the remaining ten produced more, leading one to suspect that our
enhancements may have been too high.  However, this depends entirely
on how much of the material is removed during whatever processes turn
the resulting configurations into terrestrial systems: they are quite
widely spaced.

\subsection{Evolution of inter-embryo spacing}
\label{bevol}

As noted by \cite{gold}, the argument offered in \citet{ko2} to derive
an estimate for the inter-embryo spacing $b$ as a function of $M$ and
$\Sigma$ is of dubious applicability.  Recall that the spacing $b$ in
their model is derived by equating the decrease in $b$ due to embryo
growth (due to planetesimal accretion) with an increase in $b$ due to
scattering.  The two-body scattering formula of \cite{petithenon} has
a problem in the oligarchic context: namely, there are embryos on both
sides of our scattering pair who are instead trying to scatter the two
objects closer together.  Naively one would expect this to turn the
$b$ growth from a monotonic process into a much slower random walk.
The formula also assumes that the eccentricities of the two objects
are near zero at conjunction, an assumption which breaks down in our
model as the damping timescale increases with the dissipation of the
gas.  (Simple experiments confirm both suspicions: the multiple-embryo
case shows substantially slower $b$ growth than the two-embryo case,
and introducing a decay in the strength of the eccentricity damping in
the two-embryo case also washes out the growth.)  Finally, when type I
migration is active, differential migration of equal-mass objects will
tend to increase b, and this is a non-negligible effect in our regime.

Nevertheless, our numerical experiments suggest that $b\!\simeq\!10$
is a reasonable approximation during the period we apply the
semi-analytic model, possibly accidentally: embryos quickly scatter
when $b < 5$ and growth is a weak function of $M$ ($b \propto
M^{2/15}$ in \citealt*{ko2}) so substantial masses are required before
$b=10$ becomes unacceptable.  In any case, corrections to the `wrong'
chosen b occur on short timescales, and during testing no significant
correlations between initial $b$ (granted $b\sim\!10$) and the results
were observed, suggesting that the equilibration process leaves little
signature.

\begin{figure}
\epsscale{.80}
\plotone{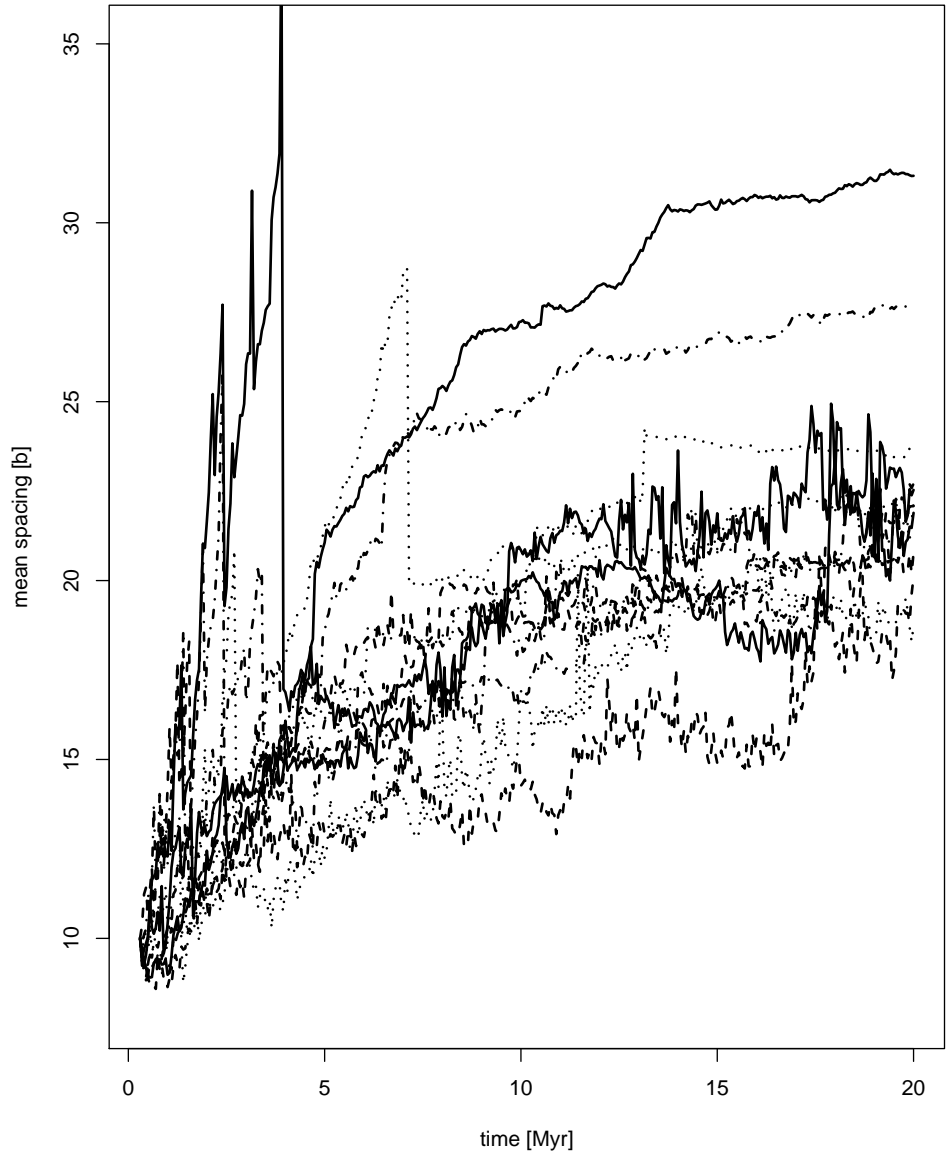}
\caption{Evolution of mean inter-embryo spacing measured in units of
single-planet Hill radii in region 0.5-1.5 AU.
\label{figure:bevol}
}
\end{figure}

Figure \ref{figure:bevol} shows the evolution of the mean
mass-weighted $b$ in the region $a=0.50-1.50\rm{AU}$ for all
simulations.  Although the specific evolution varies from integration
to integration, the trend of increase in $b$ from $\sim\!10$ to
$\sim\!20$ is clear.  The trend to large $b$ is accentuated by the
tidal migration: in our control runs without tidal migration (not
shown) we find values of $b\!\sim10-15$ were common in the late
stages.  The spikes near 4 and 6 Myr are the result of unusual
structures being formed; the 4 Myr spike in run C5 which reaches off
the scale is to be further discussed in \S\ref{walls}.  One concern is
that we find spacings of more than 20 Hill radii are typically
sufficient to produce stable systems (at least on timescales of 100
Myr, the expected formation timescale).  Preliminary simulations
suggest that the introduction of Jupiter and Saturn stirs the system
enough to produce embryo interaction, and resonance sweeping involving
the removal of the disc potential (here neglected: see \citealt{naga})
is another possibility.  We will return to this issue in a subsequent
paper.

\subsection{Material Transport}
\label{transport}

\begin{figure}
\epsscale{.80}
\plotone{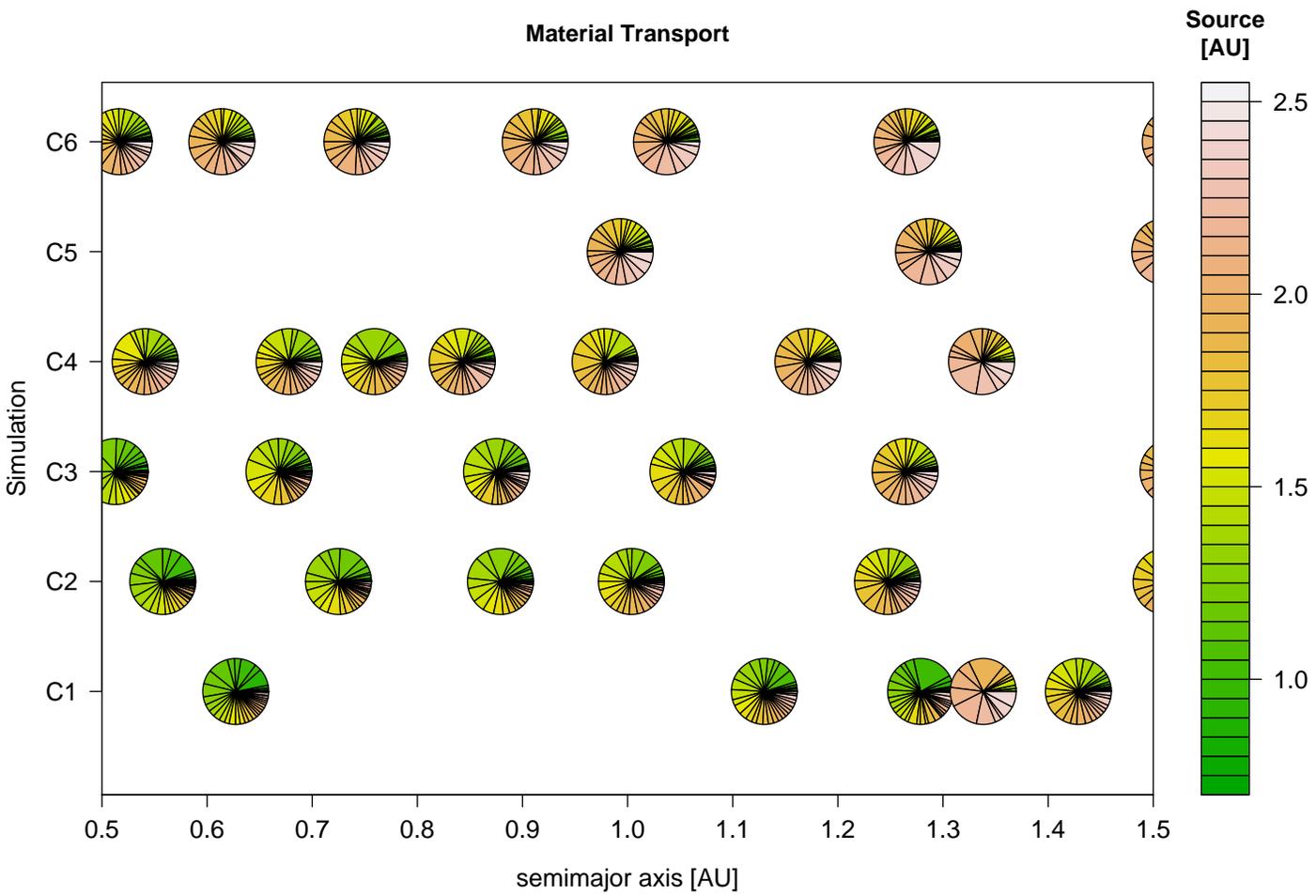}
\caption{Material fractions for C1-C6. Pies correspond to embryos at
20 Myr with slices corresponding to source regions as indicated on the legend.
\label{figure:pie1}}
\end{figure}

\begin{figure}
\epsscale{.80}
\plotone{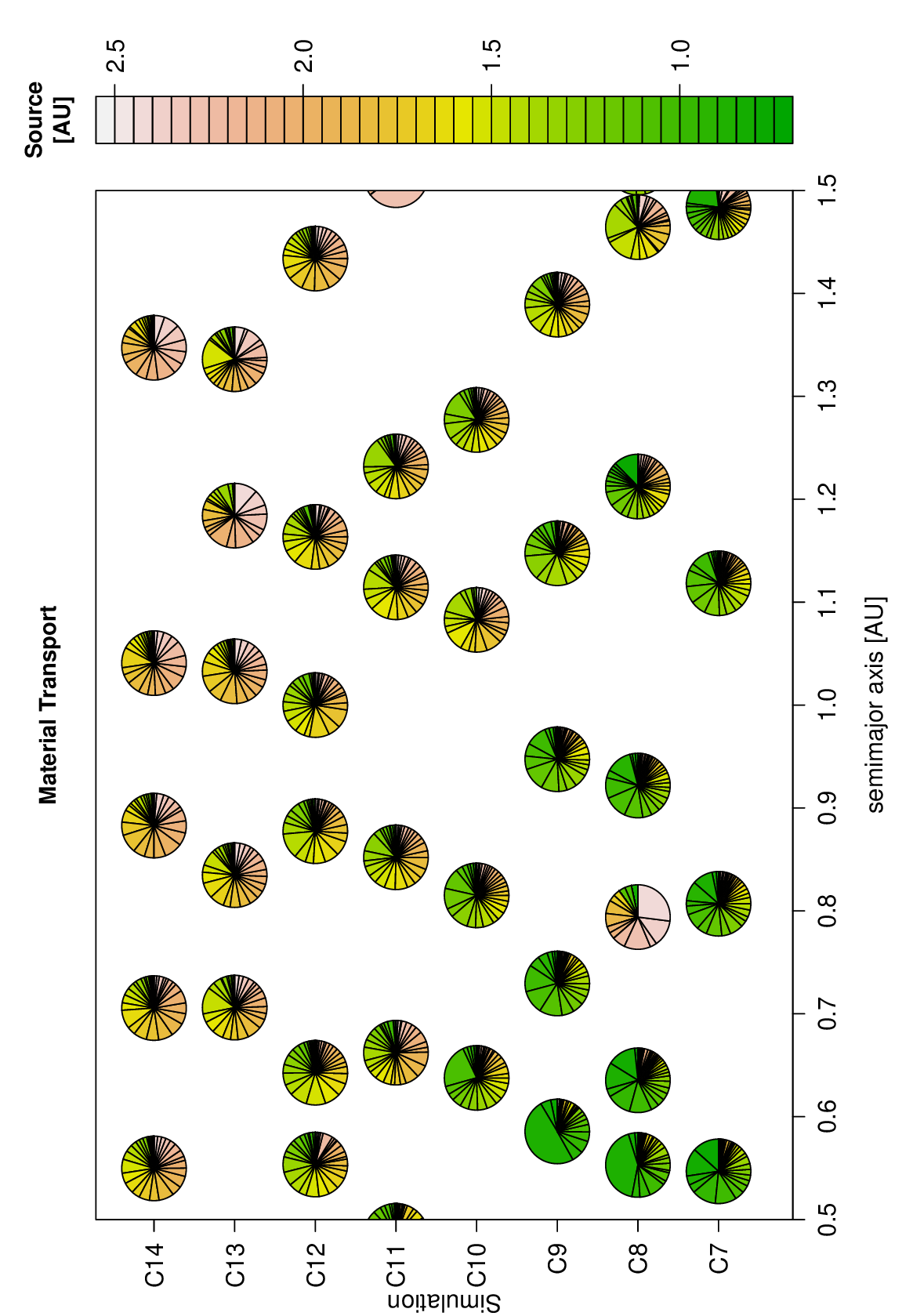}
\caption{Material fractions for C7-C14.
\label{figure:pie2}}
\end{figure}

The gas drag (both aerodynamic and tidal) tends to drive embryo and
field material inwards.  Scattering amongst embryos, or between
embryos and field objects, can counteract this and move material
outwards.  The question of how the transport processes affect the
resulting proportions of mass in the final embryos therefore arises
(i.e.~what amount of mass, from where.)  Figures \ref{figure:pie1} and
\ref{figure:pie2} show the resulting material fraction by source
region (whether embryo or planetesimal.)

Substantial amounts of the material which ends up in the 0.5-1.5 AU
region comes from beyond 1.5 AU, and even from beyond 2 AU.  In some
rare cases (e.g.~the embryo at 1.2 AU in C1, or at 0.87 AU in C8), an
embryo in fact consists mostly of material from beyond 2 AU.  It is
not surprising there are general trends with $\tau_{\rm{decay}}$.
Indeed, in fig. \ref{figure:pie1}, we see in runs C4, C5, and C6 (each
with $\tau_{\rm{decay}}$ = 2 Myr), more material from beyond 2 AU is
incorporated than in C1-C3 (with decay timescale 1 Myr.)  A somewhat
weaker trend with migration efficiency $c_a$ is also evident.
Comparatively little material originally from 1 AU becomes part of
embryos beyond 1.2 AU.  We do not believe that our lack of embryos and
planetesimals beyond 2.5 AU at $T = 0.3$ Myr plays a significant role
in these results: in the majority of the resulting embryos, material
from beyond 2.25 AU contributes little, and the arguable exception of
C6 is the simulation with the most migration (the result of a 2 Myr
dissipation timescale and an efficiency $c_a = 1$).

The large degree of inward mixing suggests that present-epoch local
chemistry may be probing much further out in the planetary disc than
might have been expected without migration, at least if countervailing
effects such as disc turbulence are limited.  Consequences of
different models of material transport during terrestrial planet
formation on presently observed water abundances are discussed in
\cite{lunine} and \cite{raymond}.

\label{section:weiden}
Recently Weidenschilling, in unpublished work, has also been
investigating terrestrial formation in enhanced discs with migration
using a hybrid dynamical and statistical technique and reached broadly
consistent conclusions (private communication, DPS 2004.)  However, he
reports a definite tendency for material to accumulate in the inner
regions.

As we remove objects when they get below 0.4 AU, and do not simulate
any objects inside of 0.75 AU at our start of 0.3 Myr, we are
insensitive to the production of a terrestrial-mass Mercury.  This
decision was motivated by our early experiments, which demonstrated
that the oligarchic predictions of the \S\ref{section:model} model
were reliable (except for certain details of the $b$ spacing) over a
large mass range before type I migration played a role and were
tolerable, although less accurate, at tidally significant masses.  The
resulting embryos were massive enough that they invariably migrated
inside of 0.4 AU before the gas vanished.  Although these early
embryos escaped our region of interest, this is not a guarantee that
they would succeed in migrating all the way to the Sun.  To
investigate the likely future of embryos whose dynamics we stopped
tracing, we can take the objects removed for falling off the inner
edge of our simulation and integrate the tidal migration equations
(\ref{eq:v_M}) and (\ref{eq:ta_semianal}).  For $\Sigma \propto
a^{-1}$, then $da/dt \propto \exp(-t/\tau_{\rm{decay}})$, and for
$\Sigma \propto a^{-3/2}$, $da/dt \propto
a^{-1/2}\exp(-t/\tau_{\rm{decay}})$.

\begin{figure}
\epsscale{.80}
\plotone{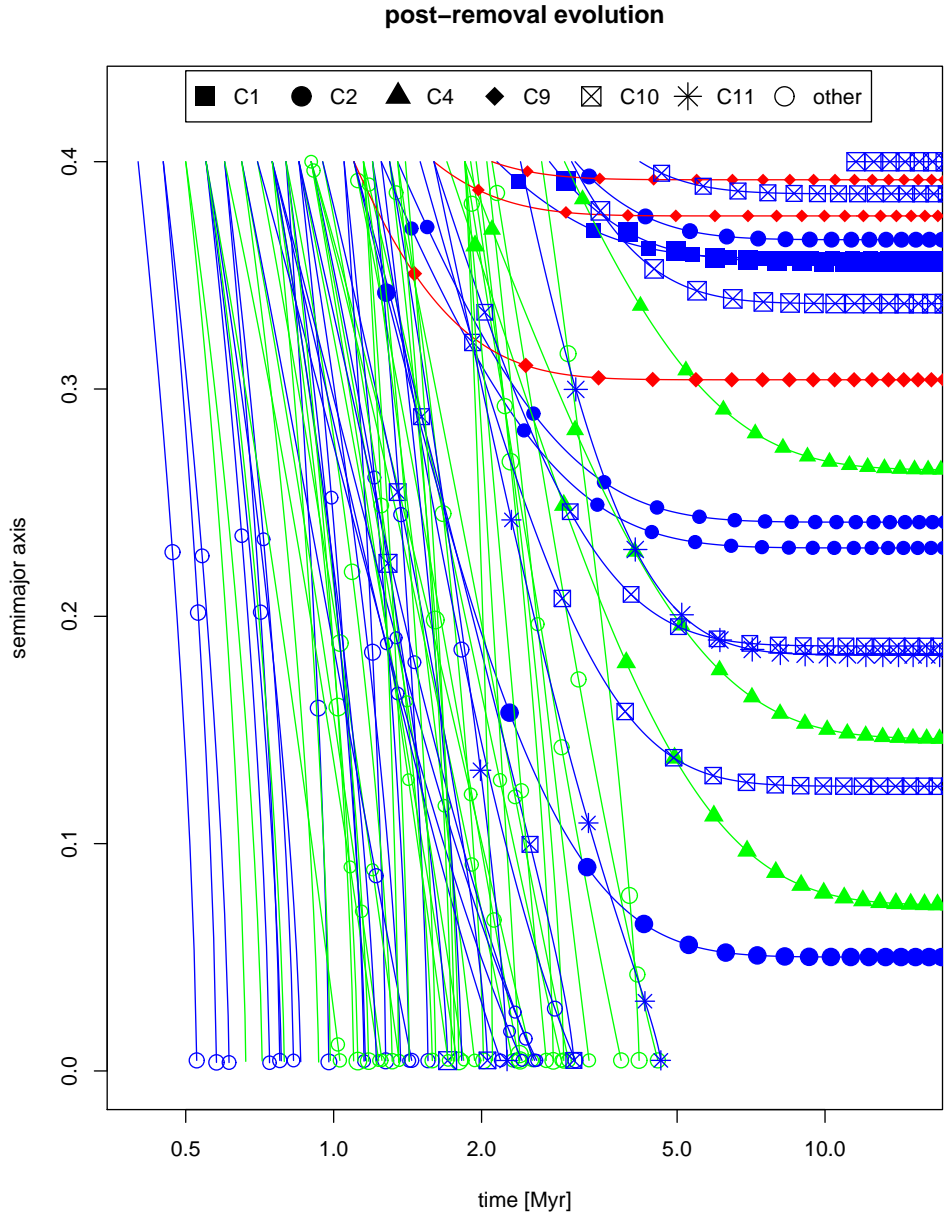}
\caption{\label{figure:wdev}
Projected future evolution of embryos removed from the
simulation.}
\end{figure}

Figure \ref{figure:wdev} shows the projected evolution of the escaped
embryos, considered one at a time.  The majority of objects (80\%)
succeed in making it to the solar radius ($\sim\!0.005$ AU), and so
our removal of them from the simulation is defensible.  In over half
of the simulations, no embryo which was removed would have survived.
In several runs (C1, C2, C4, C9, C10, C11), multiple embryos would
have survived, although in only four of these (C2, C4, C10, C11) would
an embryo have survived inside of 0.2 AU.  Note this analysis is
restricted to the embryos which were actually integrated and then
removed, not to embryos which should have been there but never were
(e.g.~embryos initially located at 0.5 AU).  A fortiori most of the
embryos we did not include inside of 0.75 AU would have also made it
to the surface of the Sun, at least for the 1 Myr and 2 Myr decay
timescales.  In the 0.5 Myr case, with its resulting lower migration,
although we are not missing much material which should end up in the
0.5-1.5 AU region, we could be missing a substantial amount between
0.2 and 0.4 AU by our choice of initial embryo range.  As for
planetesimals, given the effectiveness of migrating Jacobi wings as a
sweeping mechanism, it seems likely that many of the planetesimals not
incorporated into the interior embryos would move with the flow, and
given the strong dependence on semimajor axis of formation timescale
there are unlikely to be many left at 0.4 AU when tidally-significant
embryos are emerging at 1 AU and beyond.  In summary, for
$\tau_{\rm{decay}} \geq$ 1 Myr, we expect that we are missing only a
few embryos, and do not believe our choice of inner edge significantly
affects our conclusions.

The outer regions present their own set of difficulties: an average of
$3 M_\Earth$ of material is left beyond 1.5 AU (recall that the
initial outermost embryo in each simulation is located at $\sim2.5$
AU), which is an order of magnitude above the mass of Mars ($0.11
M_\Earth$).  As is evident from the discussion of
\S\ref{subsection:final}, we have provided an upper bound on the
necessary disc enhancement to survive type I migration, as our
enhancements consistently produce more mass than desired.  Decreasing
the enhancement to reduce this overproduction will mitigate this
problem somewhat, but even a one-third decrease will still leave $\sim
2 M_\Earth$ of material.

This problem is not unique to our migration models, however: the naive
minimum mass model, $\Sigma_{\rm{solid}} = 7.1 {\, \rm{g \, cm^{-2}}}
(r/{\rm AU})^{-1.5}$, itself produces $\sim\!1.1 M_\Earth$ of material
between 1.5 and 2.5 AU.  The majority of the unwanted material in this
region in our simulations is simply a reflection of how much was there
to begin with.  One can deal with this problem in several ways: assume
a primordial decrease in original surface density in the region
between Earth and Mars; increase the material transport rates in some
fashion, e.g.~collisional grinding producing large amounts of dust
which leaves the region very quickly due to aerodynamic drag; and so
on.  Any approach which solves the problem for non-enhanced models
without migration will have a natural analogue in our enhanced models
with migration, and therefore the problem is no worse in our model
than for the standard scenario.

Of course, it is unlikely that a simple power-law profile accurately
describes the density of the protoplanetary disc (either solids or
gas) to within a few radii of the Sun.  If the solid density reaches a
local maximum near 1 AU, it may be possible to avoid any interior and
exterior problems entirely.

\subsection{Rings}
\label{walls}
An unexpected side effect of neglecting the self-interactions of the
field particles became apparent during some of the simulations.
Occasionally, due to an early embryo-embryo merger, an object is
produced which is prematurely massive relative to its neighbours.  The
increase in mass results in a higher migration rate than the next
outer embryo, and a gap opens up between them where the only objects
are planetesimals.  This region devoid of embryos can also be created
by a major scattering event.  In a weak sense this process occurs
during every merger, but most of the time the gap is quickly filled by
another embryo, whether through migration or scattering.  However,
under certain circumstances, this gap is not closed quickly with the
entrance of an embryo from further out in the disc, but instead
persists.  In the absence of embryos there is nothing to stir the
field, and the aerodynamic drag decreases the field eccentricity,
resulting in a thin ring of planetesimals (except where an exterior
embryo arrives to form a new Jacobi wing.)  In extreme cases the width
of the ring is substantial (e.g.~0.2 AU), corresponding to a large
amount of mass ($\sim\!1.0 M_\Earth$).  At this point it can serve as a
wall, and act as a barrier to the entry of inward-migrating embryos
from beyond the ring.

There are several possibilities for the resulting behaviour.
Sometimes the ring is quickly disrupted, especially if it is small.
At other times the ring endures until an exterior embryo pushes
through.  This often leads to a surprisingly quick transition of the
embryo through the ring, a consequence of the process of forming
Jacobi wings.  The ring may or may not survive this embryo crossing.
Sometimes the ring endures, and the evolution stalls as the embryo
masses reached exterior to the ring are insufficient to push through
the barrier.

\begin{figure}
\epsscale{.755}
\plotone{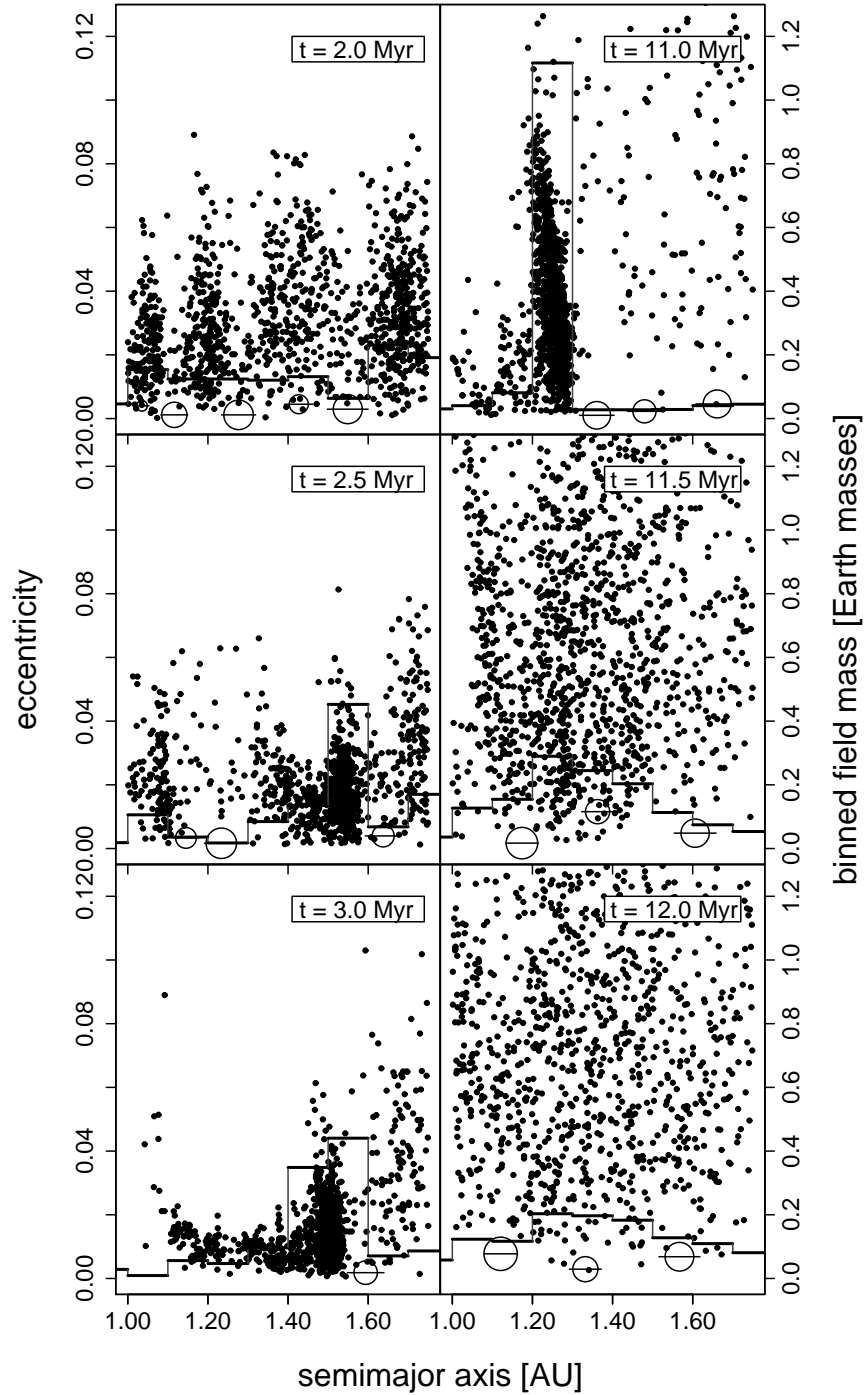}
\caption{Extreme example of the formation of a planetesimal ring.  The
ring endures from 2.5 Myr to 11.0 Myr in run C5.  The left axis shows
the eccentricity of the objects (dots) while the right axis shows the
mass of planetesimals in bins of width 0.1 AU (solid lines).
\label{figure:wall}
}
\end{figure}

An example of an extreme scenario (from run C5) is demonstrated in
figure \ref{figure:wall}, the most massive (by a factor of several)
and longest-lived (also by a factor of several) ring in our
simulations.  At 2 Myr, the system is unremarkable: between 1.0 and
1.7 AU there are four embryos separated by $\sim\!0.1$ AU, with
planetesimals throughout.  By 2.5 Myr, however, scattering events have
resulted in a region from 1.3 AU to 1.6 AU without any embryos, but
with almost an Earth mass of planetesimal material.  After another
half million years, the innermost embryos have migrated out of the
region, and only one embryo of $\sim\!0.2 M_\Earth$ remains to control
the dynamics.  (This wide embryo-free region is responsible for the
highest spike in $b$ in fig.\,\ref{figure:bevol}.)  Over the next 8
Myr, the ring moves from $\sim\!1.5$ AU to $\sim\!1.25$ AU as a result
of both aerodynamic drag (over the first several million years, at any
rate; the dissipation timescale in C5 is 2 Myr) and the more
significant push given by the exterior embryo.  During this push, the
embryo consumes enough of the ring during the Jacobi scattering to
grow from $0.2$ to $0.3 M_\Earth$.  Finally, an encounter between the
shepherd embryo and and its neighbour scatters the shepherd into the
ring by 11.5 Myr.  When the embryo enters, it quickly migrates through
and disrupts the ring.  By 12 Myr the ring is gone.

We emphasize that this is an extreme case, and that when the rings do
not form or are quickly disrupted by the embryos (before field
interactions would be expected to do so) the two-component
embryo/field approximation we make remains acceptable.


Could these rings be physical, or are they merely artifacts of our
simplified treatment of the planetesimal field?  It is true that some
early mergers and scattering events are going to occur, and
differential migration will then attempt to produce a gap.  However,
in a system with a more realistic mass spectrum (with objects of
intermediate mass) the second-tier objects which are not accreted
during the migration of the overmassive embryo might ascend to be the
new oligarchs in the absence of competition.  Furthermore, eventually
mutual planetesimal-planetesimal encounters (which we neglect) should
produce new seed objects for oligarchic growth within the ring.  Rings
which last for less than the timescale for internal disruption seem
possible, and may occur in real planetary systems.  In this case,
their chief effect would seem to be slowing the migration process for
embryos with exterior semimajor axes.

\section{Discussion}
\label{section:disc}

Various authors, looking with dismay at the unhappy consequences of
type I migration for forming giant cores, have concluded that the
effect simply does not occur.  Some recent work \citep{mg} has also
argued that the timescale for migration in some circumstances may be
substantially larger than we have assumed here.  This may ultimately
prove correct, but several points are worth making in response.
First, given the complexity of the local gas physics near the embryo,
it is probably premature to rule out near-nominal migration; a clear
understanding of the behaviour will need to wait for the next
generation of hydrodynamic simulations.  For example, it is entirely
possible that whatever effects hypothetically shut off type I
migration before the opening of a gap only become significant for
embryo masses near or beyond Earth mass.  If so, the intuition that
type I is migration is too destructive to have occurred in our system
could be completely justified regarding the giant planets but not
relevant for terrestrial formation.  Disc properties such as opacity
on which migration may sensitively depend are extremely poorly
constrained, and saying anything firm will require improved proplyd
observations.

That said, it seems possible that the gas disc is far less smooth than
we have assumed, and type I migration may be halted by abrupt changes
in gradient, opacity, or density at unknown locations.  For example,
\cite{mats} develops models for the photoevaporation of the disc which
result in a zone of depleted gas which could protect the terrestrial
region from marauding proto-Jovian embryos.  Although these specific
models have dynamical consequences which make them difficult to accept
at face value, the possibility of significant disc inhomogeneities
causing similar effects remains.  Unfortunately all of this will
clearly be very difficult to model.  Disc turbulence can wash out the
effect of tidal migration on short timescales \citep{laugh1, nelson}
but may not remove a net drift; we hope to study oligarchy in a
turbulent disc in future work.

\label{caveats}

Several compromises made to complete the integrations with the
available resources may affect the results.  The neglect of field
self-interaction contributes to the formation of rings, although to
the degree that the behaviour of the system truly is oligarchic, the
embryo-field model should be a reasonable approximation.
Nevertheless, our representation of the mass spectrum is limited, and
no field object can promote itself to an embryo.  We have chosen 100
km as the underlying planetesimal size, which is both plausible (the
formation timescale for such objects is comparable to the absolute
time when we begin the simulation) and conventional in terrestrial
formation simulations.  Smaller planetesimal particles, such as those
produced in a fragmentation spectrum, would lead to different
accretion rates \citep{raf2004} in potentially complicated ways, and
would also be likely to result in higher migration rates for
planetesimals due to aerodynamic drag, possibly requiring greater disc
enhancement.

The artificial increase of the seed mass for the embryos may also
cause problems.  It changes the gradient in embryo mass, and means
that at a given time, the outermost embryos are more massive relative
to their inner neighbours than they should be.  When the evolution is
local, this is defensible, but during migration it will lead to an
overestimation of the transport.  The overestimation should be mild in
practice as the outermost objects are sub-migratory at $T$ = 0.3 Myr,
when our N-body simulations begin.

\section{Conclusion}
\label{section:conc}

We have investigated the likely conditions at the beginning of the
terrestrial endgame as a result of type I migration during the
mid-to-late transition.  For reasonable values of the parameters we
find that plausible progenitors for the terrestrial planets can
survive despite type I migration at near the nominal rate, provided
the disc density is enhanced above the minimum model.  In particular,
in order to get the right amount of material left in the terrestrial
region with a gas dissipation timescale of $\sim\!1$ Myr, one must
work with discs several times more massive than the minimum model,
although a lower enhancement than we used here would likely suffice.
(Preliminary investigations to be reported elsewhere using discs of
$\sim2$ times the minimum mass model are encouraging, and show good
agreement with expectations.)  As a consequence, the resulting systems
at the beginning of the late phase show significant differences from
the conditions normally considered (e.g.~\citealt{cham1, cham2}): (1)
there is often a large amount, $\sim\!0.40 M_\Earth$ of material
remaining in the form of planetesimals (supporting the direction taken
in \citealt{cham2}) at the time of dissipation; (2) the embryos tend
to have larger masses, with mode $0.4 M_\Earth$; and (3) the embryos
are separated by $\sim\!20-25$ single-planet Hill radii.

In a forthcoming paper we take the next step and study what must be
done to turn our new endgame conditions into terrestrial planets.  As
our final configurations are stable when left alone, we will require
the introduction of new events (e.g.~the formation of Jupiter and
Saturn as in \citealt*{komin}) -- to complete the terrestrial
construction project.

\acknowledgments

The authors thank the sponsoring agencies of the McKenzie project, the
Canadian Institute for Theoretical Astrophysics, the Canada Foundation
for Innovation, and the Ontario Innovation Trust.  The authors are
very grateful for the generous allocations of time granted to the
project by the McKenzie external users program at CITA, without which
this research would not have been possible.  The authors acknowledge
useful discussions with Ed Thommes and Paul Wiegert.  DSM is grateful
for the assistance of Robin Humble and Chris Loken at CITA.  MJD
gratefully acknowledges ongoing support from the Natural Science and
Engineering Research Council of Canada.  HFL acknowledges support from
NASA's TPF Fundamental Research Program.

\pagebreak


\clearpage


\begin{thebibliography}{}

\bibitem[Adachi et al.(1976)]{adachi} Adachi, I., Hayashi, C., 
\& Nakazawa, K.\ 1976, Progress of Theoretical Physics, 56, 1756 

\bibitem[Artymowicz(2004)]{artym} Artymowicz, P.\ 2004, Debris Disks
and the Formation of Planets: A Symposium in Memory of Fred Gillett,
ASP Conference Series 324

\bibitem[Binney \& Tremaine(1987)]{bt} Binney, J., \& Tremaine, S.\
1987, Princeton, NJ, Princeton University Press

\bibitem[Canup et al.(2000)]{canupbook} Canup, R.~M., Righter, 
K., \& et al.\ 2000, Origin of the earth and moon, edited by R.M.~Canup and 
K.~Righter and 69 collaborating authors.~Tucson : University of Arizona 
Press ; Houston : Lunar and Planetary Institute, c2000.~ (The University of 
Arizona space science series)

\bibitem[Chambers(2001)]{cham2} Chambers, J.~E.\ 2001, 
Icarus, 152, 205 

\bibitem[Chambers \& Wetherill(1998)]{cham1} Chambers, J.~E., 
\& Wetherill, G.~W.\ 1998, Icarus, 136, 304

  
\bibitem[Duncan et al.(1998)]{dll} Duncan, M.~J., Levison, 
H.~F., \& Lee, M.~H.\ 1998, \aj, 116, 2067 

\bibitem[Fernandez \& Ip(1984)]{fernip} Fernandez, J.~A., \& 
Ip, W.-H.\ 1984, Icarus, 58, 109 
  
\bibitem[Gladman(1993)]{gladstable} Gladman, B.\ 1993, Icarus, 
106, 247 

\bibitem[Goldreich \& Tremaine(1980)]{goldtrem} Goldreich, P., 
\& Tremaine, S.\ 1980, \apj, 241, 425 

\bibitem[Goldreich et al.(2004)]{gold} Goldreich, P., 
Lithwick, Y., \& Sari, R.\ 2004, \araa, 42, 549 

\bibitem[Haisch et al.(2001)]{haisch} Haisch, K.~E., Lada, 
E.~A., \& Lada, C.~J.\ 2001, \apjl, 553, L153 

\bibitem[Hayashi(1981)]{hay1981} Hayashi, C.\ 1981, Progress of 
Theoretical Physics Supplement, 70, 35 

\bibitem[Ida \& Makino(1992a)]{ida92a} Ida, S., \& Makino, J.\ 
1992, Icarus, 96, 107 

\bibitem[Ida \& Makino(1992b)]{ida92b} Ida, S., \& Makino, J.\ 
1992, Icarus, 98, 28 

\bibitem[Ida \& Makino(1993)]{ida93} Ida, S., \& Makino, J.\ 
1993, Icarus, 106, 210 

\bibitem[Kinoshita et al.(1991)]{kyn} Kinoshita, H., 
Yoshida, H., \& Nakai, H.\ 1991, Celestial Mechanics and Dynamical 
Astronomy, 50, 59 

\bibitem[Kokubo \& Ida(1996)]{ko1} Kokubo, E., \& Ida, S.\ 
1996, Icarus, 123, 180 

\bibitem[Kokubo \& Ida(1998)]{ko2} Kokubo, E., \& Ida, S.\ 
1998, Icarus, 131, 171 

\bibitem[Kokubo \& Ida(2000)]{ko2000} Kokubo, E., \& Ida, S.\ 
2000, Icarus, 143, 15 

\bibitem[Kokubo \& Ida(2002)]{ko_div} Kokubo, E., \& Ida, S.\ 
2002, \apj, 581, 666 

\bibitem[Kominami \& Ida(2004)]{komin} Kominami, J., \& Ida, 
S.\ 2004, Icarus, 167, 231 


\bibitem[Laughlin et al.(2004)]{laugh1} Laughlin, G., 
Steinacker, A., \& Adams, F.~C.\ 2004, \apj, 608, 489 

\bibitem[Lunine et al.(2003)]{lunine} Lunine, J.~I.,
Chambers, J., Morbidelli, A., \& Leshin, L.~A.\ 2003, Icarus, 165, 1

\bibitem[Masset \& Papaloizou(2003)]{masset} Masset, F.~S., \&
Papaloizou, J.~C.~B., \apj, 588, 494

\bibitem[Matsuyama et al.(2003)]{mats} Matsuyama, I., 
Johnstone, D., \& Murray, N.\ 2003, \apjl, 585, L143 

\bibitem[Menou \& Goodman(2004)]{mg} Menou, K., \& Goodman, J.\ 2004,
\apj, 606, 520


\bibitem[Nagasawa et al.(2003)]{naga} Nagasawa, M., Lin, 
D.~N.~C., \& Ida, S.\ 2003, \apj, 586, 1374 

\bibitem[Nelson \& Papaloizou(2004)]{nelson} Nelson, R.~P., \& 
Papaloizou, J.~C.~B.\ 2004, Astronomical Society of the Pacific Conference 
Series, 321, 367 

\bibitem[Papaloizou \& Larwood(2000)]{paplar} Papaloizou, J.~C.~B., \&
Larwood, J.~D.\ 2000, \mnras, 315, 823

\bibitem[Papaloizou \& Lin(1984)]{papandlin} Papaloizou, J., \& 
Lin, D.~N.~C.\ 1984, \apj, 285, 818 

\bibitem[Peale \& Lee(2002)]{peale} Peale, S.~J., \& Lee, 
M.~H.\ 2002, Science, 298, 593 

\bibitem[Petit \& Henon(1986)]{petithenon} Petit, J.-M., \& Henon, M.\
1986, Icarus, 66, 536

\bibitem[Rafikov(2004)]{raf2004} Rafikov, R.~R.\ 2004, \aj, 128, 1348 

\bibitem[Raymond et al.(2004)]{raymond} Raymond, S.~N., Quinn, 
T., \& Lunine, J.~I.\ 2004, Icarus, 168, 1 

\bibitem[Stewart \& Ida(2000)]{si} Stewart, G.~R., \& Ida, S.\ 2000,
Icarus, 143, 28


\bibitem[Tanaka \& Ida(1997)]{tanaka} Tanaka, H., \& Ida, S.\ 
1997, Icarus, 125, 302 

\bibitem[Tanaka et al.(2002)]{tanakaward} Tanaka, H., Takeuchi, 
T., \& Ward, W.~R.\ 2002, \apj, 565, 1257 
 
\bibitem[Thommes(2001)]{edphd} Thommes, E.~W.\ 2001, Ph.D.~Thesis

\bibitem[Thommes et al.(2003)]{edolig} Thommes, E.~W., Duncan, 
M.~J., \& Levison, H.~F.\ 2003, Icarus, 161, 431 

\bibitem[Ward(1986)]{ward} Ward, W.~R.\ 1986, Icarus, 67, 164

\bibitem[Wetherill \& Stewart(1989)]{weth89} Wetherill, G.~W., 
\& Stewart, G.~R.\ 1989, Icarus, 77, 330 

\bibitem[Wetherill \& Stewart(1993)]{weth93} Wetherill, G.~W., 
\& Stewart, G.~R.\ 1993, Icarus, 106, 190 

\bibitem[Wisdom \& Holman(1991)]{wishol} Wisdom, J., \& 
Holman, M.\ 1991, \aj, 102, 1528 


\end{thebibliography}
\end{document}